\documentclass[a4paper,10pt]{article}

\usepackage{a4wide}
\usepackage[utf8]{inputenc}
\usepackage{blindtext}
\usepackage{hyperref}
\usepackage{fancyhdr}

\usepackage{amssymb, amsfonts, amsthm}

\usepackage{mathbbol}

\usepackage{authblk}

\usepackage{placeins}
\usepackage{mathtools}
\usepackage[toc,page]{appendix}

\usepackage{bm}
\usepackage{natbib}
\usepackage{multicol}
\usepackage[dvipsnames]{xcolor}
\usepackage{caption, booktabs}
\usepackage{setspace}

\theoremstyle{definition}

\newfont{\handw}{cmmi10 scaled 800}
\newfont{\handws}{cmmi10 scaled 600}

\providecommand{\keywords}[1]{\noindent\textbf{\textit{Keywords}}: #1}

\newcommand\commentout[1]{}

\usepackage{tabularx}
\usepackage{subcaption}
\numberwithin{equation}{section}

\newcommand{\nil}[1]{}

\title{Asymmetry Analysis of Bilateral Shapes} 
\author{Kanti V. Mardia$^{\text{1,}**}$, Xiangyu Wu$^{\text{1,}*}$, John T. Kent$^{\text{1,}*}$, Colin R. Goodall$^{\text{1,}*}$, \\Balvinder S. Khambay$^{\text{2,}*}$}
\affil{{\footnotesize $^{\text{\sf 1}}$Department of Statistics, School of Mathematics, University of Leeds, LS2 9JT, England.}}
\affil{{\footnotesize $^{\text{\sf 2}}$Third Orthodontics Division, Dentistry, School of Health Sciences, College of Medicine and Health, University of Birmingham, England.}}
\affil{{\footnotesize $^{\text{**}}$Email: k.v.mardia@leeds.ac.uk.}}

\begin{document}

\setstretch{2}
\maketitle

%\journaltitle{Journals of the Royal Statistical Society}
%\DOI{DOI HERE}
%\copyrightyear{XXXX}
%\pubyear{XXXX}
%\access{Advance Access Publication Date: Day Month Year}
%\appnotes{Original article}

%\firstpage{1}

%\author[1,$\ast$]{Kanti V. Mardia}
%\author[1]{Xiangyu Wu}
%\author[1]{John T. Kent}
%\author[2]{Colin R. Goodall}
%\author[3]{Balvinder S. Khambay\ORCID{0000-0000-0000-0000}}

%\authormark{Mardia et al.}

%\address[1]{\orgdiv{Department of Statistics}, \orgname{University of Leeds}, \orgaddress{\country{United Kingdom}}}
%\address[2]{\orgdiv{Department}, \orgname{Organization}, \orgaddress{\street{Street}, \postcode{Postcode}, \state{State}, %\country{Country}}}
%\address[3]{\orgdiv{Department}, \orgname{University of Birmingham}, \orgaddress{\country{United Kingdom}}}

%\corresp[$\ast$]{Corresponding author. \href{email:k.v.mardia@leeds.ac.uk}{k.v.mardia@leeds.ac.uk}}

%\received{Date}{0}{Year}
%\revised{Date}{0}{Year}
%\accepted{Date}{0}{Year}
  
\begin{abstract}
  Many biological objects possess bilateral symmetry with some noise about a midline or midplane.
  This paper uses landmark-based methods to measure departures from bilateral symmetry,
  especially for paired data and two-group problems where one group may be more symmetric than the other.
  We formulate our work under size-and-shape analysis framework including pre-registration under rigid body motion.
  We then construct a vector of elementary asymmetry features defined per landmark coordinate per object.
  Next, we introduce two approaches for comparing bilateral symmetry between the two groups or paired data:
  (1) the elementary features are combined into a scalar asymmetry measure for each object,
  then standard univariate tests are used to test for asymmetries;
  (2) a univariate test statistic is constructed for each feature, then
  the maximum of these statistics leads to an overall test statistic and important features
  can be extracted from the data. Our methodology is illustrated on two registered datasets collected 
  to assess the success of two kinds of surgeries: (1) cleft lip surgery data, where the symmetry
  is compared between a group of cleft lip subjects and normal subjects;
  (2) orthognathic surgery data, where the patients’ symmetry after the orthognathic surgery
  are compared to that before surgery.
\end{abstract}

\keywords{asymmetry measures, cleft lip surgery, orthogonathic surgery, registration, shape analysis, size-and-shape analysis of smile.}

%\maketitle

\section{Introduction}\label{sec:intro}
Bilateral symmetry, also called left-right symmetry, is a key property in many biological settings.  An
object is said to be
bilaterally symmetric if its reflection through some midline (for objects in two dimensions) or
midplane (for objects in three dimensions) passing through the center of 
the object is exactly the same as the original object.  In practice,
exact bilateral symmetry seldom holds and it is of interest to study
the extent of any asymmetry.  Research on the symmetries of objects
in the real world has a long history \citep[see, for example,][]{weyl1952}. Statistical
methods to test asymmetry were first formalized in
\citet{kvm2000}. Subsequently, a wide variety of statistical methods
have been developed; see, for examples, \citet{kvm2001},  \citet{bock2006},
\citet{ajmera2022} and \citet{ajmera2023}.

The focus of this paper is on landmark-based objects as in
\citet{kvm2000}.  Each object is a set of $K$ labelled points or
landmarks, represented as a $K \times M$ configuration matrix giving
the positions of $K$ landmarks in $M$ dimensions.  The cases $M=2$ and
$M=3$ are the most important in practice. We call an object
\emph{bilateral} if its landmarks can be divided into two categories:
\emph{pairs} and \emph{solos}. 
An object is \textit{bilaterally symmetric} about an $M-1$-dimensional midplane $\mathcal{P}$
if all the solos lie on the midplane, and the two landmarks in each pair are equally spaced
about $\mathcal{P}$, meaning that their midpoint lies on $\mathcal{P}$ and the vectors to 
each from their midpoint are orthogonal to $\mathcal{P}$.
Let there be $K_P$ pairs and $K_S$ solos; then $K=2K_P+K_S$. 
%Registration of objects ensures that the midplane $\mathcal{P}$ has the same coordinate axis or axes, which for left-right symmetry the first coordinate is commonly used. 
After we have determined the $M$ orthogonal axes for the object, the intersection point between the midplane and the first coordinate axis can be computed. Under translation and rotation it is ensured that the midplane $\mathcal{P}$ passes through the origin and is orthogonal to the first coordinate axis.
All objects in the dataset should be registered so that all of them share
the same midplane $\mathcal{P}$, hence, further analyses are available.
If the dataset has not been pre-registered before the analyses,
then registration is required; the details are given in Section \ref{sec:asym:reg}.
% where paired landmarks locate on both sides of $\mathcal{P}$, while solos are unpaired
%If the
%dataset has not been pre-registered, then registration is required.
%The details of registration are given in Section \ref{sec:asym:reg}. 
\nil{(\textbf{\textcolor{red}{The order of these para might need to change. Either we talk about registration and asymmetry measurements after introducing the two data, or we first mention the registration, asymmetry measurements at first, then introduce the data.}})}
%\textcolor{red}{our composite score extends the asymmetry measures used by
%\citet{bock2006} and \citet{BV2023} (\textbf{move to Sec 3.1})}.

This paper is motivated by two medical questions related to the level of asymmetry in a person's smile.
One is related to cleft lip surgery which is performed so that
form and function can be restored to particularly lip symmetry at rest and during movement.
%the smile can be normal after surgery
%and in the years following.
The second is related to orthognathic surgery, which
%. For cleft lip patients,  surgery 
%is performed so that 
%form and function can be restored particularly lip symmetry at rest and during movement.
%the smile can be normal after surgery.
%Orthognathic surgery 
is performed to correct jaw discrepancies in order to restore the form and function but the surgery can influence facial and lip symmetry.
%their smile.

We have been provided with appropriate data in each case. We will refer the cleft lip data as Smile Data 1,
and the orthognathic surgery data as Smile Data 2.
In both datasets, we have a set of landmarks on the lip area as appropriate to each case; 
 Smile Data 1 contains 24 landmarks over the lip while Smile Data 2 contains 16 landmarks over the lip
and nose. More details of the data are provided in Section \ref{sec:smile} and Section \ref{sec:jaw} respectively.
In each case, the subjects were asked to smile from rest (close lip) to maximum open lip smile and a
four-dimensional (three spatial dimensions
plus time) movie capture was made of the face using Di4D motion capture system (\url{https://di4d.com}).
A superimposed finite element mesh combined with optical flow techniques were used to follow the landmarks through the frames of the capture, taken at one-second intervals, of the capture. Note that these include both anatomical landmarks, e.g. the corners of the mouth, and pseudo landmarks, e.g. a midpoint of the upper right lip defined at first frame and then followed through successive frames of the capture by optical flow.
%These landmarks are 
%followed through the frames of the capture.

Smile Data 1 contains two
 groups: normal subjects and subjects who had previous cleft lip surgery. 
Smile Data 2 includes a single group of subjects before and after orthognathic surgery,
so it is a paired dataset.
 The aim is to assess the success of these two different surgeries,
 that is, are the cleft lip subjects more symmetric after the surgery,
and has orthognathic surgery had an impact on symmetry.
 Both datasets are pre-registered in the same way and we have three fixed frames
 (the beginning, middle and end of the smile) for each subject.
%and the orthognathic subjects

%We end the paper with some discussion in Section \ref{sec:discussion}.
The asymmetry information in an object is represented as a vector
of \emph{coordinatewise elementary asymmetry features}, $\bm d$, say,
defined in Section \ref{sec:asym:elem}.  Increasing
the magnitude of any element in $\bm d$ indicates greater
asymmetry. The elementary feature vector is also the starting point
of \citet{bock2006} and \citet{BV2023}, and we define its elements explicitly  
in Section \ref{sec:asym:elem}.

Given two groups of objects, a natural question is whether one group
is more asymmetric than the other.  Section \ref{sec:test} looks at
two testing strategies which we name combine-then-compare and compare-then-combine respectively. 

Our methodology involves simultaneous comparisons, hence, meta-analysis
is considered.
The literature on meta-analysis as required in the paper is scattered over several papers,
and in Section \ref{sec:meta-analysis}, we give an overview of meta-analysis as required for our work.
%Section gives the details of appropriate meta-analysis for our asymmetry analysis,
The meta-analysis is carried out in Section
\ref{sec:smile} and \ref{sec:jaw} respectively for the two datasets.
%since .

%In Section \ref{sec:smile}, we  analyze  Smile Data 1, and 
%in Section \ref{sec:jaw}, Smile Data 2 is analyzed. 
The paper concludes with a discussion
in Section \ref{sec:discussion}. 
%In Section \ref{sec:meta-analysis}, another two methods for hypotheses
%testing are introduced and performed on the two datasets.
%\textcolor{red}{The procedures for estimating the midline or midplane are given in 
%Supplementary.}

\section{Describing asymmetry}\label{sec:asym}

\subsection{Registration}\label{sec:asym:reg}
We introduce the following notations: let $X$, with elements $X[k,m]$, be a $K \times M$ configuration
matrix giving the positions of $K$ landmarks in $M$ dimensions for a
single object or subject or individual, where $k=1,\dots,K$ and $m=1,\dots,M$. We write as $X[k, \    ] \in \mathbb{R}^M$ 
for the coordinates of the $k$th landmark, $k=1,\dots,K$. The most important choice
is $M=3$, and this case is emphasized in the presentation, but the
mathematical theory is valid for any dimension $M \geq 1$.  In general, upper
case letters are used for matrices and bold lower case letters for
vectors.  In addition the letters $K,M,J$ are reserved for the number
of landmarks, number of dimensions, and number of elementary features
(Section \ref{sec:asym:elem}), respectively,
with typical indices given by the lower case letters $k,m,j$.  In
general, each data object $X$ will be viewed as a noisy version of a
bilaterally symmetric configuration $X^S$.

{\bf Rigid body transformations} If $X$ is a configuration, then a \emph{rigid body motion} takes $X$
to $X^* = \bm 1_M \bm c^T + X R$,  say, where $\bm c \in \mathbb{R}^M$ is a
translation vector and $R$ is an $M \times M$ rotation
matrix.  For the purposes of this paper, we say that $X$ and $X^*$
have the same \emph{size-and-shape}, as the size-and-shape of $X$ is the
equivalence class of configurations under the group of rigid body
motions (see, for example, \citet{drydenmardia2016}).
%following the standard terminology in shape analysis 
%.  That is,  
The asymmetry information carried by the size-and-shape is of interest and 
such information is unaffected under rigid body motion. The size 
is emphasized here, since the size information is taken into consideration.  
Recall that the term ``shape'' is a standard terminology in statistical shape analysis which
means the equivalence class under the larger
group of similarity transformations so the scale is also filtered out.
%which also allows for a change in size.  

A hyperplane $\mathcal{P}$ in $\mathbb{R}^M$ can be written in the
form $\mathcal{P}= \{\bm x \in \mathbb{R}^M : \bm n^T \bm x = b \}$, in
terms of a unit normal vector $\bm n$, say, and a scalar offset term
$b$, say.  When $M=3$ a hyperplane becomes a two-dimensional plane,
and when $M=2$ a hyperplane becomes a one-dimensional line.  For
simplicity we refer to hyperplanes as planes everywhere.  Under a
rigid body motion, $\bm n$ is transformed to $R^T \bm n$ and $b$ is
transformed to $b + \bm n^T R \bm c$.

{\bf Two types of Registration} 
We distinguish two types of registrations
\begin{enumerate}
    \item[(a)] Axis registration.
    \item[(b)] Basis registration.
    %\textcolor{blue}{where all the bases of the $M-1$-dimensional midplane $\mathcal{P}$ are identified, together with the axis orthogonal to $\mathcal{P}$}.
\end{enumerate}

Suppose there is an associated midplane $\mathcal{P}$ for every
bilateral data object $X$.  Ways in which $\mathcal{P}$
might be determined are discussed below. For data analysis, it is
helpful to ``register'' $X$ using a rigid body motion so that after
transformation $\bm n = \bm e_1$ and $b=0$, i.e., the midplane passes
through the origin and the normal direction is given by the first
coordinate axis, while still keeps the size-and-shape of $X$ unchanged.

%We will distinguish two types of registration:
% In this work, we only use basis registration, so the details of axis registration are given in Section 2 of Supplementary.

\begin{figure}[!h]
    \centering
    \includegraphics[scale=0.35]{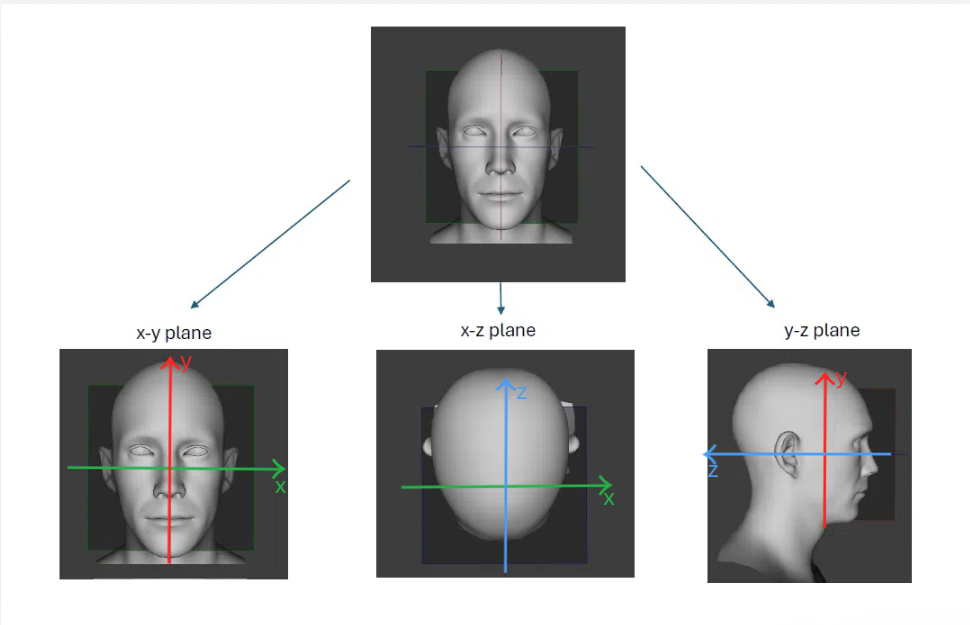}
    \captionsetup{width=1.0\textwidth}
    \caption{Coordinates system used for registration of human face used in our illustrative smile examples: with three principal planes. In the figure at top row, the $x$-$y$ plane is in green, the $y$-$z$ plane is in red and the $x$-$z$ plane is in blue. For the three sub-figures in the bottom row, the $x$-axis is in green, $y$-axis is in red and $z$-axis is in blue.}
    \label{axes}
\end{figure}

In some settings it is possible to carry out a stronger version of
registration than the axis registration, we will call \emph{basis registration}, in which all the
coordinate axes have a natural interpretation. In our illustrative smile examples, for 3D
measurements of the human head, the midplane is known as the sagittal
plane, and it is natural to require the coordinate axes to have the
following interpretations:
\begin{itemize}
    \item[(a)] $x$-coordinate: left-right 
    \item[(b)] $y$-coordinate: down-up
    \item[(c)] $z$-coordinate: back-front
\end{itemize}
Figure \ref{axes} shows these
coordinate axes as well as the three planes on a human face which have been used in pre-registration for our smile data. 
That is, the $x$-axis is positioned horizontally on the 
face (positive direction is left from the view point of the subject), the $y$-axis is placed vertically on 
the face (positive direction is upward), whereas the $z$-axis is positioned 
in the direction inwards and outwards of the face (positive direction is outward), see Figure \ref{axes}. 
%Further details are given in Section \ref{sec:smile}.

%The coordinates given in Section \ref{sec:asym:reg} can be
%related to these coordinate axes as follows:\\
%$x$-coordinate: left-right \\
%$y$-coordinate: down-up\\
%$z$-coordinate: back-front.

In general, to carry out basis registration, it is also necessary to know or
estimate $M-1$ meaningful basis vectors in
$\mathcal{P}$.  
\nil{By ``hybrid'', we mean
that the following procedure is used:
(i) the normal direction to the midplane is estimated by
combining an object with its reflection and using generalized
Procrustes analysis (GPA), (ii) a symmetric ``average shape'' is found from all the
objects using GPA, (iii) the midplane for
the average shape is oriented so the basis directions have natural
interpretations using expert knowledge, and (iv) each data
configuration is aligned to the average shape using ordinary
Procrustes analysis (OPA). The details are given in \citet{drydenmardia2016}.
\textcolor{blue}{Note that the differences between the Procrustes methods used here
and in axis registration are that we neither rotate the midplane $\mathcal{P}$ for the
average shape nor we use OPA to align each object to the average shape for axis registration.}}
In this case the three possibilities become: (a) the basis is known a priori;
(b) the basis is estimated using expert knowledge; or 
(c) the basis is estimated using hybrid Procrustes analysis.
For details of Procrustes analysis,
see, for example, \citet{goodall1991} and \citet{drydenmardia2016}. 
%The details of basis registration are contained in the Section 2 of Supplementary.

For our illustrative smile examples,
%smile data applications in Section \ref{sec:smile} and \ref{sec:jaw}, 
expert knowledge (scenario (b)) has been used to estimate the midplane with the
stated interpretations for the coordinate axes as above (Figure \ref{axes}).
It may be noted that
there will be one more reflection before the registration in some data; 
specifically, for those cleft lip subjects in Smile Data 1 with the cleft on the right,
we reflect in the midplane so that the cleft is consistently on the left.

\subsection{Elementary features}\label{sec:asym:elem}

Suppose the size-and-shape dataset has been basis registered about a midplane
$\mathcal{P}$ that passes through the origin and is normal to the
first coordinate axis and the $M-1$ axes of $\mathcal{P}$ are all determined.
Then it is possible to define various
elementary features to describe the asymmetry of each data object.
Paired and solo landmarks are treated separately.
We begin with an idealized bilaterally symmetric object $X^S$.  Each of the $K$ landmarks of $X^S$ will be either to the left of $\mathcal{P}$, to the right of $\mathcal{P}$, or on $\mathcal{P}$. The counts of landmarks in these subsets are respectively $K_L$,  $K_R$, and $K_S$  where  $K_L = K_R = K_P$ .

%Let $K_L$ and $K_R$ denote the total number of landmarks on the left
%or right of $\mathcal{P}$. We have $K_L=K_R=K_P$.
For any $K \times M$ observation $X$ modeled on $X^S$, landmarks, identified by their indices, follow the prescription of $X^S$.
%\textbf{\textcolor{red}{The numbers of landmarks to the left or right of, or on the estimated midplane can certainly differ from $K_L$, $K_R$, $K_S$.}}
Let $k_L$ and $k_R$ denote the left and right indices for a typical
landmark pair from a bilateral object $X$ and consider the $M$
\emph{coordinatewise signed elementary features} for the landmark pairs,
\begin{align}
  d[(k_L,k_R),1] &= X[k_L,1]+X[k_R,1] \label{eq:pair1}\\
  d[(k_L,k_R),m] &= X[k_L,m]-X[k_R,m], \quad m=2, \ldots, M. \label{eq:pair2}
\end{align}
That is, there is one feature for each coordinate of each landmark
pair, so there are in total $MK_P$ coordinatewise features.
The first coordinates for the two landmarks in
one pair have opposite signs, while other coordinates have the same
sign with respect to $\mathcal{P}$. Note that this pattern
is true for both axis and basis registrations. For $X^S$, the quantities in equations (\ref{eq:pair1})
and (\ref{eq:pair2}) should all be 0. Thus, the quantities given in (\ref{eq:pair1}) and (\ref{eq:pair2})
quantify the departure from symmetry.
%taking the sum of the first coordinates and differences for other coordinates
Similarly, if $k_S$ is a typical solo landmark, consider the single
elementary feature 
\begin{equation}
  \label{eq:solo}
  d[(k_S)] = X[k_S,1].
\end{equation}
There are in total $K_S$ elementary features for solos. For $X^S$, the first
coordinate of a solo landmark will be 0, with no 
restriction on other coordinates. Hence, a single feature as in
(\ref{eq:solo}) is adequate to describe the asymmetry of a solo landmark. 

The coordinatewise elementary features can be collected into a
\emph{signed elementary feature vector} $\bm d = (d_j)$, say, of
length $J_{\text{basis}} = MK_P + K_S$.  The elements of $\bm d$ will either
be listed sequentially with a subscript index, i.e.,
$d_j, \ j=1, \ldots, J_{\text{basis}}$, or using parentheses within
square brackets, as in (\ref{eq:pair1})-(\ref{eq:solo}).
Note that the vector $\bm d$ is a function of matrix $X$ and $\mathcal{P}$.

When only axis registration is available, the coordinate information
in (\ref{eq:pair2}) does not have a well-defined
interpretation. This is because the coordinates except the first one 
of paired landmarks would be changed after rotation within $\mathcal{P}$, 
hence leads to changes in values of (\ref{eq:pair2}) even though the asymmetry 
information does not change (but note that equation (\ref{eq:pair1}) does not change). 
Further, at each landmark the coordinate information
can be collected into a single number, called a \emph{landmark
  elementary feature} which is defined per landmark,
\begin{equation}
  \label{eq:pair3}
  d^*[(k_L,k_R)] = \left\{ \sum_{m=1}^M d^2[(k_L,k_R),m] \right\}^{1/2}, \  d^*[(k_S)] = |X[k_S,1]|,
\end{equation}
where $d$'s are defined by (\ref{eq:pair1}) and (\ref{eq:pair2}). 

Another useful way to think about the elementary features is by using
reflection.
%Let $X$ be a basis registered configuration, and 
Let $X^{(\text{refl})}$ denote the reflection of $X$ about $\mathcal{P}$.
Computationally, this matrix is obtained by changing the sign of the
first column of $X$ and interchanging the row indices for each
landmark pair. Then the elementary features of $X$ can be obtained by
comparing the elements of $X$ and $X^{(\text{refl})}$.  If $X$ is
bilaterally symmetric about $\mathcal{P}$, then $X =X^{(\text{refl})}$ and $\bm d = \bm 0$.

In many examples, such as the illustrative smile examples
studied later in this paper in Section \ref{sec:smile} and \ref{sec:jaw},
the direction of asymmetry, for example, left or right
asymmetry, can vary between individuals and is not thought
interesting. We will focus on the absolute values of
the elementary features.  Define the \emph{absolute elementary feature vector}
$\bm a$ with elements
\begin{equation}
  a_j= |d_j|, \quad j=1, \ldots, J,
  \label{absd}
\end{equation}
where $J = J_\text{basis}$ or $J = J_\text{axis}$, as appropriate. In other words, 
only the extent of the asymmetry is interesting.
The elements of $\bm a$ can also be written 
in a similar way as elements of $\bm d$ in (\ref{eq:pair1})-(\ref{eq:pair3}).
In the rest of the paper, each configuration $X$ is reduced to an
absolute elementary feature vector $\bm a$ for further numerical and
statistical analysis. Note that $\bm a$ is also a function of matrix $X$ and $\mathcal{P}$.  

\subsection{Example 1}\label{sec:asym:eg}
We give here a simple example to illustrate our basic notation and construction of the vector $\bm a$ and related quantities.
Consider two configurations in $M=2$ dimensions with $K=4$ landmarks given by
\begin{align}
    \begin{split}
        \label{eq:squareeg-configs}
 X_1 = \begin{pmatrix}
     X_1[1, \   ]^T \\ X_1[2, \   ]^T \\ X_1[3, \   ]^T \\ X_1[4, \   ]^T
 \end{pmatrix} = \begin{pmatrix} -1 & \phantom{-}0 \\ \phantom{-}0 & \phantom{-}1 \\
 \phantom{-}1 &     \phantom{-}0 \\ \phantom{-}0 & -1 \end{pmatrix}, \  \   
 X_2 = \begin{pmatrix}
     X_2[1, \   ]^T \\ X_2[2, \   ]^T \\ X_2[3, \   ]^T \\ X_2[4, \   ]^T
 \end{pmatrix} = \begin{pmatrix} -0.95 & \phantom{-}0.36 \\ -0.28 & \phantom{-}2.11 \\
 \phantom{-}0.99 & \phantom{-}0.54 \\ -0.31 & -1.37 \end{pmatrix}.
    \end{split}
\end{align}

\begin{figure}[!h]
    \centering
      \begin{subfigure}[!h]{0.15\textwidth}
         %\centering
         \includegraphics[scale = 0.45]{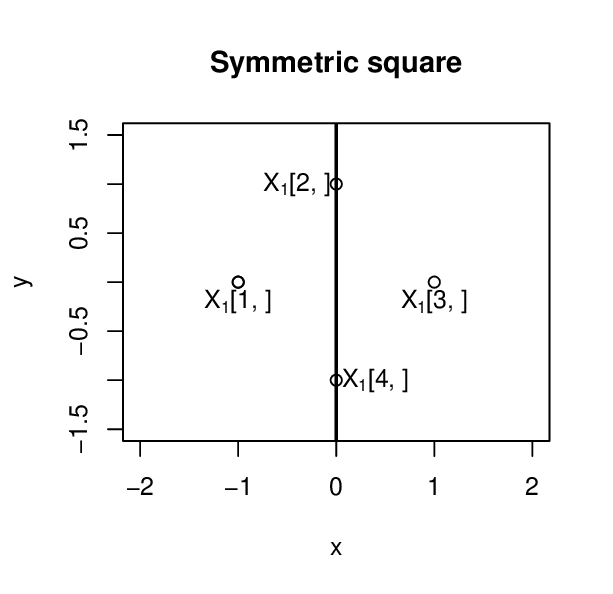}
         %\subcaption{Symmetric square $X_1$.}
         %\label{sym_sq}
     \end{subfigure}
     \hspace{20mm}
     %\vspace{-3mm}
     %\hskip 1em
     %\bigskip
     %\hfill
     \begin{subfigure}[!h]{0.15\textwidth}
         %\centering
         \includegraphics[scale = 0.45]{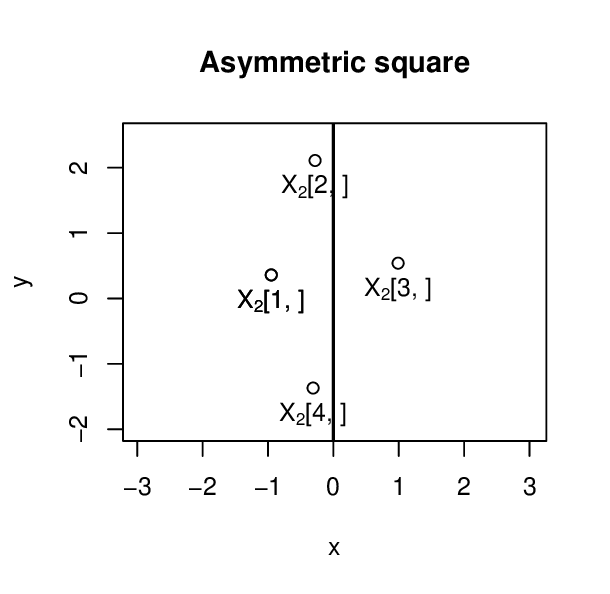}
         %\subcaption{Asymmetric square $X_2$.}
         %\label{asy_sq}
     \end{subfigure}
     %\hfill
     %\hskip 1em
     \vfill
     \vspace{-3mm}
     \begin{subfigure}[!h]{0.15\textwidth}
         %\centering
         \includegraphics[scale = 0.45]{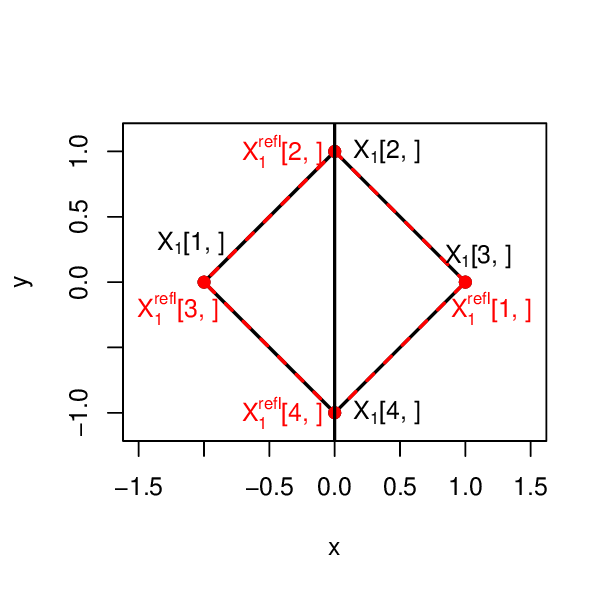}
         \captionsetup{width=1.2\textwidth}
         %\subcaption{$X_1$ and $X_1^{\text{(refl)}}$}
         %\label{sym_sq_ref}
     \end{subfigure}
     %\bigskip
     \hspace{20mm}
     %\hskip 1em
     %\hfill
     %\vspace{-8mm}
     \begin{subfigure}[!h]{0.15\textwidth}
         %\centering
         \includegraphics[scale = 0.45]{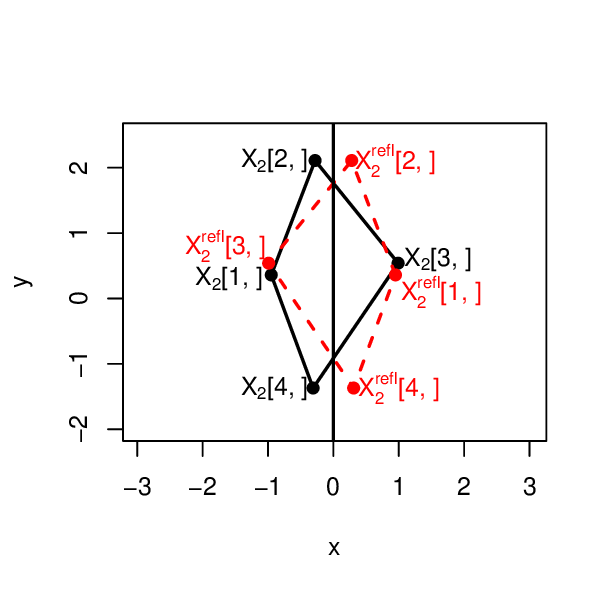}
         \captionsetup{width=1.2\textwidth}
         %\subcaption{$X_2$ and $X_2^{\text{(refl)}}$}
         %\label{asy_sq_ref}
     \end{subfigure}
     %\vspace{3mm}
     \captionsetup{width=1.0\textwidth}
  \caption{The left and right figures in the top row show the original configurations of symmetric square $X_1$ and asymmetric square $X_2$ (in black) respectively. The left and right figures in the bottom row show the original objects (in black solid lines) together with their respectively reflections $X_1^{\text{(refl)}}$ and $X_2^{\text{(refl)}}$ (in red dotted lines) respectively, in order to illustrate the coordinatewise elementary feature vector.}
  \label{fig:X1X2}
  \end{figure} 
%coordinatewise elementary feature vector

%Suppose the landmarks consist of one landmark pair,
Therefore, there is one landmark pair
$(k_L, k_R)=(1, 3)$ and two solos, $k_S=2, 4$, so $K_P=1$ and
$K_S=2$.  Further, suppose both configurations are treated as basis
registered with $\mathcal{P}$ normal to the first coordinate axis, i.e. $\mathcal{P}$ is the $y$-axis.

The configurations $X_1$ and $X_2$ are shown in the left and right figures
in the top row of Figure
\ref{fig:X1X2} respectively, whereas the bottom row of Figure
\ref{fig:X1X2} show their corresponding reflections, where the left
is for $X_1$ and $X_1^{\text{(refl)}}$ and right is for $X_2$ and $X_2^{\text{(refl)}}$.
Note that $X_1$ is bilaterally
symmetric, with $\bm a = \bm 0$.  Table \ref{squareeg-features} gives
the $J_\text{basis} = 4$ elements of the vector $\bm a$ for $X_2$.
The  two elements for the landmark pair are listed first, and the two
elements for the solos listed last, with
$$\bm a = (0.04,\ 0.18,\ 0.28,\ 0.31)^T$$ for $X_2$ .  It can be seen
visually in Figure \ref{fig:X1X2} that for $X_2$, landmarks 2 and 4 deviate
more from the $y$-axis than the landmark pair (1, 3) differs from
symmetry. Hence the last two components of $\bm a$ are larger than the
first two components.

  \begin{table}[!t]
    \centering
        \captionsetup{width=1.0\textwidth}
    \caption{Example in Section \ref{sec:asym:eg}.  Elements of the absolute elementary feature vector
    $\bm a \in \mathbb{R}^4$ for $X_2$ with $K=4$ landmarks in $M=2$ dimensions.}
    \label{squareeg-features}
    %\vspace{3mm}
    \begin{tabular}{|c|c|c|c|}
    \hline 
    Landmark Indices & Coordinate Axis & Feature of $\bm a$ & value of $\bm a$\\ \hline
   Pair (1,3)       &   1        & $a[(1,3),1]=|X_2[1,1]+X_2[3,1]|$ &  0.04\\ \hline
   Pair (1,3)       &   2        & $a[(1,3),2]=|X_2[1,2]-X_2[3,2]|$ &  0.18\\ \hline
   Solo 2           &   1        & $a[(2)]=|X_2[2,1]|$       &  0.28   \\ \hline
   Solo 4           &   1        & $a[(4)]=|X_2[4,1]|$       &  0.31 \\ \hline
    \end{tabular}
\end{table}

\section{Hypothesis tests}\label{sec:test}
Consider a dataset of $N$ basis registered configurations observed on two
groups of individuals, where the first $N_1$ configurations belong to
one group (Group 1) and the last $N_2$ configurations belong to the other group (Group 2),
$N = N_1+N_2$, where it is thought that the first group of
configurations might be more symmetric than the second. One way to
think about this question is to construct a one-sided hypothesis test
of 
\begin{align}
    \begin{split}
        &H_0: \text{ the two groups have the same distribution of asymmetry vs.}\\
        &H_1: \text{ Group 1 is more symmetric than Group 2.}
    \end{split}
\label{originH0H1}
\end{align} 
For the paired data, we change the Group 1 and Group 2 in the above hypotheses
to experimental and control groups respectively, where same subjects are contained
in the two paired groups.

We will give two general approaches to construct a test statistic in
order to compare the symmetries between the two groups: (a)
combine-then-compare, where univariate test methods are used,
and (b) compare-then-combine, where we form the hypotheses under union
intersection test framework.  These two
approaches will be explored in the next subsections.

%\textcolor{red}{To specify the null and alternative hypotheses explicitly, some
%notations are needed. (\textbf{can be removed})}
Suppose the data take the form of $N$ registered
configurations
\begin{equation}
  \label{eq:data}
  X_n, \ n=1, \ldots, N.
  \end{equation}
  Let $\bm a_n$ denote the corresponding absolute elementary feature
  vector for $X_n$ (equation (\ref{absd})), with elements
  $a_{nj}, \ j=1, \ldots, J$ and $J=J_{\text{basis}}$ for the rest of this paper.

  Then the hypotheses given in equation (\ref{originH0H1}) can be
  reformulated as:
\begin{align}
    \begin{split}
        H_0: &\text{ $\bm a_n \sim F$ i.i.d. for all $n$ vs.} \\
        H_1: &\text{ $\bm a_n \sim F_1$ i.i.d. for $n=1,\dots,N_1$, $\bm a_n \sim F_2$ i.i.d. for $n=N_1+1,\dots,N$,}
    \end{split}
\end{align}
where $F$, $F_1$, $F_2$ are multivariate cumulative density functions (c.d.f.s) and $F_1$ is
stochastically smaller than $F_2$ (that is,
$F_1(\bm a) \geq F_2(\bm a)$ for all $\bm a$ with at least one strict
inequality, see, for example \citet{stoyan1983}). In other words, the distribution of $F_2$ 
is shifted towards right comparing with $F_1$, which indicates more asymmetries.
Under the paired data setting, $\bm a_n$ for $n=1,\dots,N_1$ represent the experimental
data and $\bm a_n$ for $n=N_1+1,\dots,N$ represent the control data, where $N=2N_1$.

%[This final paragraph to be rewritten once Discussion sorted]\\
We concentrate in this section on t-test and Mann-Whitney U test. In Section \ref{sec:smile}, we further discuss
some alternative univariate tests as well as multivariate tests in relation to the application of Smile Data 1.

\subsection{Testing strategy: combine-then-compare}
\label{sec:test:score}

%Let $\bm a$ denote an absolute elementary feature vector.  
Each element $a_j$ of $\bm a$ (equation (\ref{absd})) is nonnegative and a larger value of $a_j$ indicates
greater asymmetry.  Hence it is natural to reduce the $J$-dimensional
vector $\bm a$ to a single number by combining the elements of $\bm a$
into a \emph{composite asymmetry score}
\begin{equation}
  \label{eq:cas}
  u = \phi(\bm a),
\end{equation}
say.  Here the function $\phi(\bm a)$ is assumed to be 
%\textcolor{red}{monotonically
%increasing in the sense that it is (\textbf{redundant})}
monotonically increasing in each
component $a_j$ when the other elements are held fixed.

Let $\psi(a)$ be a monotonically nonegative increasing function of a scalar
argument $a \geq 0$.  Then one way to define $\phi(\bm a)$ is as an
\emph{additive composite asymmetry score}
\begin{equation}
  \phi_{\psi, \bm w}(\bm a)  = \sum_{j=1}^{J} w_j \psi(a_j),
\label{eq:acas}
\end{equation}
where the $w_j\geq 0$ are pre-assigned weights. 
Some choices for weights
include the following:
\begin{itemize}
\item[(a)] equal weights $w_j = 1$, 
\item[(b)] greater weights near the midplane or away from the midplane, or
%\item[(c)] greater weights away from the midline, or
\item[(c)] different weights for pairs and solos.
\end{itemize}
For example, in case (b), a weight might
depend on $|X[k_R,1]-X[k_L,1]|$ for a landmark pair, with solos treated
separately. The choices of weights in general depend on prior belief of the extent of 
asymmetry information carried by each landmark, for example, if it is known a priori that landmarks near
$\mathcal{P}$ can carry more important asymmetry information than landmarks far away from $\mathcal{P}$, then
greater weights should be assigned to the nearer landmarks. For Smile Data 1, this information is available
so greater weights will be given near the midplane, see Section \ref{sec:smile}.

%The choice of function $\psi$ governs the relative influence of different
%magnitudes of an elementary feature.
Another way to give some elementary features greater prominence than others
is through the choice of the functions $\psi$.  
For example, the quadratic function
$\psi(a) = a^2$ is more sensitive to outlying features than the linear
function $\psi(a) = a$.

We write below explicitly
two such scores with equal weights and give them names $L_1$ and $L_2$ statistics respectively.
\begin{equation}
    \text{$L_1$ statistic: } \phi_{L_1}(\bm a) = \sum_{j=1}^J a_j=\sum_{(k_L, k_R)} \sum_{m=1}^M |d[(k_L, k_R), m]| + \sum_{k_S} |d[(k_S)]|,
\label{eq:L1score}
\end{equation}
\begin{equation}
   \text{$L_2$ statistic: } \phi_{L_2}(\bm a) = \sum_{j=1}^J a_j^2=\sum_{(k_L, k_R)} \sum_{m=1}^M d[(k_L, k_R), m]^2 + \sum_{k_S} d[(k_S)]^2,
\label{eq:L2score}
\end{equation}
where the $d$'s are given in (\ref{eq:pair1})-(\ref{eq:solo}) and $\bm a$, $a_j$ are given in (\ref{absd}).
%In the above equations, we use the two expressions for elements in $\bm d$
%and $\bm a$ introduced in Section \ref{sec:asym:elem}.

We note that \citet{bock2006} have proposed an asymmetry score which is proportional to $\phi_{L_2}$, that is 
%with weights $w_j=1$
%for landmark pairs and weights $w_j=2$ for solos and $\psi(a_j)=a_j^2$ in equation (\ref{eq:acas}), namely,
\begin{equation}
    %\sum_{(k_L, k_R)} \sum_{m=1}^M d[(k_L, k_R), m]^2 + \sum_{k_S} d[(k_S)]^2.
    \frac{2}{K} \phi_{L_2} (\bm a).
\label{eq:bock2006}
\end{equation}
Further, \citet{BV2023} uses the score given by
%The choice $\psi(a) = a^2$ was used in \citet{bock2006} with 

\begin{equation}
    \phi_{L_1}^*(\bm a) = \sum_{(k_L, k_R)}d^*[(k_L,k_R)] + \sum_{k_S} d^*[(k_S)],
\label{as1}
\end{equation}
%with weights equal to 1 for both pairs and solos.
The $d^*[(k_L,k_R)]$ and $d^*[(k_S)]$ are defined
in equation (\ref{eq:pair3}). Also, by squaring each element in the two sums in (\ref{as1}) and using (\ref{eq:pair3}), we recover $\phi_{L_2}$:
%we can define a similar score as follows using $L_2$ distances
\begin{equation}
 \phi_{L_2}^*(\bm a) = \sum_{(k_L, k_R)}d^*[(k_L,k_R)]^2 + \sum_{k_S}|d^*[(k_S)]|^2=\phi_{L_2}(\bm a).
\label{eq:as2}
\end{equation}
%It can be shown using (\ref{eq:pair3}) that this equation (\ref{eq:as2}) reduces to (\ref{eq:L2score}). That is,  we have
%$$\phi^*_{L_2}(\bm a)=\phi_{L_2}(\bm a).$$
Note that we can generalize
$\phi^*(\bm a)$ in the same way as $\phi(\bm a)$ with weights.
Instead of 
considering asymmetries with respect to each landmark as in (\ref{as1}),
(\ref{eq:L1score}) and (\ref{eq:L2score}) view all features in $\bm a$ as a whole and 
take the $L_1$ and $L_2$ norms. For the subjects which are more symmetric, 
their corresponding $\bm a_n$ should be close to the origin. 
Hence, the $L_1$ and $L_2$ distances between $\bm a_n$ with $\bm 0 \in \mathbb{R}^J$ 
are expected to be smaller for control subjects.

%\textbf{Example 2.3 continued.}
%which would have the above quantities for $X_2$ as much details as possible Patel and our $L_1$ $L_2$ such as value of $d^*$.

\textbf{Example 1 continued.} Recall the 
asymmetric square $X_2$ defined in equation (\ref{eq:squareeg-configs}).
The configuration is shown in Figure \ref{fig:X1X2}. The landmark elementary
feature, $d^*[(1, 3)]$, defined in equation (\ref{eq:pair3}) is computed and
reported in Table \ref{asX2}. The composite asymmetry scores $\phi_{L_1}(\bm a)$ (\ref{eq:L1score}),
scaled $\phi^*_{L_1}(\bm a)$ (\ref{as1}) (divided by the number of landmarks) 
and $\phi_{L_2}(\bm a)$ (\ref{eq:L2score})
are computed on $X_2$ for illustration. The results are shown in Table \ref{asX2}.
Note that the values of 
\begin{align*}
    \begin{split}
        d^*[(1, 3)] = \sqrt{(X_2[1, 1] + X_2[3, 1])^2 + (X_2[1, 2] - X_2[3, 2])^2}=0.18
    \end{split}
\end{align*}
and $a[(2)]=0.28$, $a[(4)]=0.31$ (Table \ref{squareeg-features}) 
are less than 1, so taking squares of them in $\phi_{L_2}$ leads to smaller values.
Note that here we have scaled $\phi^*_{L_1}$ by a factor of half, so unscaled $\phi^*_{L_1}=0.78$
which is similar to the value of $\phi_{L_1}=0.81$.
%The value of (\ref{eq:acas}) is less than (\ref{as1}) since squared distances are
%computed and we have from Table \ref{squareeg-features} that these distances are all
%less than 1. Thus, squared distances would lead to 

\begin{table}[t!]
    \centering
    \captionsetup{width=1.0\textwidth}
    \caption{The two composite asymmetry scores defined in equation (\ref{eq:acas}) and equation (\ref{as1}) are computed on the configuration $X_2$.}
    \label{asX2}
    \begin{tabular}{|c|c|c|c|c|}
    \hline
         & $d^*[(1, 3)]$ (\ref{eq:pair3}) & $\phi_{L_1}(\bm a)$ (\ref{eq:L1score}) & $\phi^*_{L_1}(\bm a)$ (\ref{as1}) & $\phi_{L_2}(\bm a)$ (\ref{eq:L2score}) \\
         \hline
        $X_2$ & 0.18 & 0.81 & 0.39 & 0.21 \\
         \hline
    \end{tabular}
\end{table}
%\;

Once a composite asymmetry score $\phi(\bm a)$ has been chosen, let
\begin{equation}
  \label{eq:composite-scores}
  u_n = \phi(\bm a_n), \quad n=1, \ldots, N,
\end{equation}
denote the composite asymmetry scores for the data in (\ref{eq:data}).

Suppose $u_n$ for $n=1,\ldots,N_1$ and $n=N_1+1,\ldots,N$ are realizations from 
random variables $U_1$ and $U_2$ respectively, and let $g=1, 2$ denote
the control and cleft groups. Let
\begin{equation}
  \label{eq:moms}
  \mu_g=E\{U_g\}, \  \sigma^2_g= \text{var} \{U_g\}
\end{equation}
denote the expectation and variance for a random variable
$U_g \sim F_g$ (should not be confused with the distribution of the vector $\bm a$ in Section \ref{sec:test}).
%, where $g=1,2$ labels one of the two groups. 
A simpler version of hypotheses in (\ref{originH0H1}) is
\begin{equation}
    H_0: \mu_1 = \mu_2 \text{ vs. } H_1: \mu_1 < \mu_2
\label{simpleH0H1}
\end{equation}
%Under , $\mu_1 < \mu_2$.
There are several test statistics that can be used for testing 
hypotheses given in (\ref{originH0H1}) and (\ref{simpleH0H1}),
including the following:
\begin{itemize}
\item[(a)] Parametric test: the standard two-sample t-test, which assumes $\sigma_1^2=\sigma_2^2$;
%a common variance for the two groups ();
\item[(b)] Non-parametric test: the Mann-Whitney U test, also known as the Wilcoxon rank
  sum test \citep{mann1947}.
 \end{itemize}
%The Mann-Whitney U test is nonparametric so that
%the test results are invariant under monotonic transformations of the $u_n$.
In case (a), the significance of the test statistic under $H_0$
 can be computed either analytically (assuming normality) or using a
 bootstrap approximation.  In case (b) the significance can
 be derived from combinatorial arguments. Mann-Whitney U test 
 is considered where the normality assumption on composite asymmetry scores 
 $u_n = \phi(\bm a_n)$ may fail to hold. 

\textbf{Paired data} The test strategies described above can be applied to the 
paired data by making necessary modification, namely we need to use the paired test
in place of the two-sample test for both the parametric and non-parametric cases.
Specifically, paired t-test and Wilcoxon signed-rank test are used for the parametric
and non-parametric cases respectively.
%For non-parametric case, we use the   in place of the Mann-Whitney U test.

 An advantage of the combine-then-compare approach is that it enables
 us to accumulate evidence from different features, so if there is a
 small amount of asymmetry on many different elementary features, then
 the composite score will have a large value.  Further, this approach
 enables the two groups to be compared visually with stem and leaf
 plots and in particular, any overlap between the two groups can be
 easily assessed. Applications of this approach are in Section 
 \ref{sec:combine-compare-Data1}, \ref{sec:combine-compare-Data2}
 and \ref{sec:combine-compare-Data1-2}.

 \subsection{Testing strategy: compare-then-combine}
 \label{sec:test:uit}

 In this approach, separate test statistics are constructed for each feature in
 the elementary feature vectors.  The most extreme of these separate
 test statistics is then used as an overall test statistic. Before we 
 give the details, the hypotheses in equation (\ref{originH0H1}) need to be 
 formulated in the Union Intersection Test (UIT) style:
 %(as explained in Section \ref{sec:intro}) as following:
 \begin{equation}
     H_0=\bigcap_{j=1}^J H_0^j \text{ vs } H_1=\bigcup_{j=1}^J H_1^j,
 \label{eq:UIT}
 \end{equation}
 where $H_0^j$: the mean of $j$th unsigned elementary feature between 
 the two groups is the same, whereas $H_1^j$ is that the mean of group 2 
 is larger than mean of group 1, for $j=1,\dots,J$.
 
 For each choice $j$ of an unsigned elementary feature, we construct a statistic
 to compare the two groups, for example, a two-sample t-statistic or a Mann-Whitney
 U statistic. Denote the resulting statistics by
 $v_j, \ j=1, \ldots, J$.  For example, if a t-statistic is used, then
\begin{equation}
\label{eq:v-stat}
v_j = (\bar a_j^{(1)} - \bar a_j^{(2)})/s_j.
\end{equation}
where $\bar a_j^{(g)}, \ g=1,2$ are the sample means of the $j$th unsigned
elementary feature in the two groups and $s^2_j$ is the corresponding pooled within-group
variance.

Following \citet{boyett1977}, an overall UIT statistic to test $H_0$ vs $H_1$ in equation (\ref{eq:UIT}) can be defined by taking the
 maximum of these separate statistics,
 \begin{equation}
   \label{eq:max}
   V = \max_{j=1, \ldots, J} v_j,
 \end{equation}
 and \citet{boyett1977} compute the critical points of $V$ under the null hypothesis using bootstrap.
 The landmarks are correlated with each other, 
 so the $d_j$ are not independent for $j=1,\ldots,J$ here, so we cannot
 derive the distribution of $V$ theoretically. Hence, the bootstrap has been
 used for estimating the critical points of $V$. 
 %Note that the
 %sign of $V$ indicates which group is more asymmetric.

If there exists $v_j$ which exceeds the critical value of $V$ at a pre-determined
significance level, 
then we not only know that the corresponding $H_0^j$ is rejected (hence the overall
$H_0$ in (\ref{eq:UIT}) is also rejected), but also know that 
the landmark corresponding to such $v_j$ is important.
 %[XW: it would be a good idea to explain why a bootstrap
 %is needed here, e.g. because of correlations between the variables.]

Our main aim is somewhat different from \citet{boyett1977}: 
we aim at discovering important landmarks while  \citet{boyett1977}
focuses on estimating a $p$-value. Also, the two bootstrap procedures 
are slightly different. Their procedure is as follows:
%The bootstrap procedure
%used by us is also different from their, which is:
%who have used:
\begin{enumerate}
    \item[(a)] Construct a set $E_n$ containing all random samples of size $n$ taken without replacement from origin dataset $\{\bm a_i\}_{i=1}^N$.
    %permutations of the dataset.
    \item[(b)] Sample with replacement $B$ times from this set $E_n$ to form the bootstrap resampled dataset.
    %, i.e. sample permutations of the data. The permuted samples .
    \item[(c)] Compute the test statistic $\{V^{(1)}, \dots, V^{(B)}\}$ on the resampled dataset.
    \item[(d)] Compare the test statistic $V$ computed on first $n$ observed data $\{\bm a_i\}_{i=1}^n$ to $\{V^{(1)}, \dots, V^{(B)}\}$.
\end{enumerate}
The procedure given in \citet{boyett1977} does not perform bootstrap directly on
the data, and only part of the data is used in either $E_n$ or the observed statistic,
which is probably to reduce the computational budget. In contrast, 
%we perform On the other hand, 
since our data size is not large and the computational budget is far more than enough,
we directly sample with replacement from the whole dataset with $n=N$ and use all data to compute
the observed statistic. Using all data is very likely to improve the final result.
Our procedure is as follows:
\begin{enumerate}
    \item[(a)] Sample $N$ samples with replacement from $\{\bm a_i\}_{i=1}^N$ and repeat this step $B$ times.
    \item[(b)] Compute $\{V^{(1)}, \dots, V^{(B)}\}$ on resampled dataset.
    \item[(c)] Compare the test statistic $V$ computed on all observed data $\{\bm a_i\}_{i=1}^N$ to $\{V^{(1)}, \dots, V^{(B)}\}$.
\end{enumerate}

For compare-then-combine approach, we use a similar selection procedure to Tukey's
 method \citep{tukey1949} in ANOVA though in ANOVA a separate test statistic is
 used for each pairwise difference between main effects.
%An advantage of this compare-then-combine approach versus combine-and-compare is that if the
% null hypothesis is rejected, then simultaneous confidence intervals
% can be constructed to identify those features on which the two groups
% significantly differ.  
Here, a separate test statistic for each $j$ is more appropriate since 
the different features in the unsigned elementary feature vector $\bm a$ 
measure different features of the face. An illustration is given below
in Section \ref{sec:compare-combine-Data1}
and \ref{sec:compare-combine-Data1-2}.
%for the Smile Data 1.

\textbf{Paired data} The approach given in this section can be extended to the
paired data scenario with the following changes applied to the bootstrap procedure.
%The bootstrap process now becomes:
\begin{enumerate}
    \item[(a)] Compute $a^{(2)}_{nj} - a^{(1)}_{nj}$ for each subject $n$ and each feature $j$, where $a^{(1)}_{nj}$ and $a^{(2)}_{nj}$ are the feature $j$ of vectors $\bm a_n^{(1)}$ and $\bm a_n^{(2)}$ for subject $n$ from Group 1 and 2 respectively.
    \item[(b)] Assign $+1$ or $-1$ to each $\bm a^{(2)}_{n} - \bm a^{(1)}_{n}$ randomly with equal probability, which forms the permuted sample. Repeat this step $B$ times.
    \item[(c)] Compute $t$-values on the permuted sample and record the largest $t$-value.
\end{enumerate}
An illustration of this compare-then-combine approach is given below for the Smile Data 2
in Section \ref{sec:compare-combine-Data2}.

\section{Meta-Analysis Methods}\label{sec:meta-analysis}
We can consider the meta-analysis, which carries out simultaneous inference
and also belongs to the compare-then-combine
approach category. In particular in meta-analysis, we can obtain an overall $p$-value
by combining the $p$-values from separate tests. We now give an overview for meta-analysis.
%\subsection{Methodology Details}
%In the compare-then-combine approach, we  carry out several tests and an alternative approach is to  through meta-analysis of which we  Note that 

Suppose we have $N$ basis registered configurations, where the first $N_1$ configurations
are from Group 1 and the rest $N_2$ configurations are from Group 2. 
Absolute elementary feature vectors $\bm a_1, \dots, \bm a_N \in \mathbb{R}^J$ defined in equation (\ref{absd})
can be constructed on each configuration. Let $\bm a^{(g)}$ be the random vector associated
to the population of group $g$ and let $a_j^{(g)}$ denote feature $j$ of vector
$\bm a^{(g)}$, where $g=1, 2$ for Group 1 and Group 2 respectively and $j=1,\dots,J$.
%Suppose we are interested in the parameter $\bm \beta \in \mathbb{R}^J$.
%Let $\hat{\bm \beta}$ denote the
%estimation of $\bm \beta$, which is a random variable. 
%In our applications, $\bm \beta$ will be the vector of absolute elementary
%features defined in . Consequently, $\hat{\bm \beta}$
%will be the ,

Consider the $J$ individual
tests with corresponding hypotheses
%\[H_{0,j}: \beta_j = c, \  H_{L,j}: \beta_j < c, \  H_{R,j}: \beta_j > c, \ H_{1,j}: \beta_j \neq c, \  j=1,\dots,J,\]
%\[ \]
\begin{align*}
    \begin{split}
        H_{0}^j: a_j^{(1)} = a_j^{(2)}, \   \ H_{1,L}^j: a_j^{(1)} < a_j^{(2)}, \ \   
        H_{1,R}^j: a_j^{(1)} > a_j^{(2)}, \ \   H_{1}^j: a_j^{(1)} \neq a_j^{(2)},
    \end{split}
\end{align*}
for $j=1,\dots,J$.
The $p$-values $\Tilde{p}_j$ and $1 - \Tilde{p}_j$ correspond to
the left and right alternative hypotheses $H_{1,L}^j$ and $H_{1,R}^j$ respectively.  
The two-sided $p$-value $p_j=2\min (\Tilde{p}_1, 1-\Tilde{p}_j)$ corresponds to
the two-sided alternative hypothesis $H_{1}^j$.
The way of computing $\Tilde{p}_j$ and $p_j$ depends on the choice of the univariate test.
%Let $\hat{\beta}_j^{\text{obs}}$
%denote the estimation of $\beta_j$ based on observations. Let
%\begin{equation}
 %   \Tilde{p}_j = \text{Pr}(\hat{\beta}_j \leq \hat{\beta}_j^{\text{obs}}|H_{0,j})
  %  \label{eq:pval-individual}
%\end{equation}
%denote the left-sided $p$-value under null hypothesis for test $j$ which is
%used to test $H_{0,j}$ against $H_{L,j}$. The right-sided $p$-value corresponding
%to $H_{R,j}$ is $1-\Tilde{p}_j$. The two-sided $p$-value related to $H_{1,j}$ is
%$p_j=2\min (\Tilde{p}_1, 1-\Tilde{p}_j)$

One of such methodology to combine individual $p$-values is the well-known Fisher's method 
\citep{fisher1932}. The Fisher's combination function is given as:
\begin{equation}
    P_L = -2\sum_{j=1}^J \log(\Tilde{p}_j), \   P_R = -2\sum_{j=1}^J \log(1-\Tilde{p}_j), P = -2\sum_{j=1}^J \log(p_j),
    \label{eq:test-Fisher}
\end{equation}
which correspond to the left, right and two-sided overall alternative hypotheses.
$P_L$, $P_R$ and $P$ all follow $\chi^2_{2J}$-distribution. Hence, a final $p$-value
is obtained by comparing them with the tail of the $\chi^2_{2J}$-distribution.

For the Fisher's method, $P_L$ is used in our illustrative smile examples but not $P$ or $P_R$,
since we would like to test whether the values of absolute elementary features
from Group 1 are smaller than the feature values from
Group 2. In other words, we are interested in the left-sided
tests. However, the right-sided and two-sided tests could be relevant for some other practical examples.
%so $P$ and $P_L$ are not applicable

There is another combination function based on \citet{pearson1934} which is given as
\begin{equation}
    Q = \max(P_L, P_R),
    \label{eq:test-Pearson}
\end{equation}
which does not follow a $\chi^2$-distribution. Hence, randomization tests and permutation
tests are used to estimate the overall $p$-value in two-sample and paired data cases
respectively. For the randomization tests, the procedures are given as below:
\begin{enumerate}
    \item[(a)] Sample without replacement from the pooled dataset of the two groups for $B$ times.
    \item[(b)] Perform individual two-sample tests on the resampled data and obtain $p$-values $\Tilde{p}_j^{(b)}$ and $1-\Tilde{p}_j^{(b)}$, for $j=1,\dots,J$ and $b=1,\dots,B$.
    \item[(c)] Compute the overall $p$-values defined in equation (\ref{eq:test-Fisher}) and hence $Q$ defined in (\ref{eq:test-Pearson}).
\end{enumerate}

The procedures for permutation tests are given as the following:
\begin{enumerate}
    \item[(a)]  Shuffle the subjects randomly, such as change the subjects between experimental group to control group while keeping the paired structure. Repeat this step for $B$ times.
    \item[(b)] Perform individual paired tests to obtain $p$-values $\Tilde{p}_j^{(b)}$ and $1-\Tilde{p}_j^{(b)}$, for $j=1,\dots,J$ and $b=1,\dots,B$.
    \item[(c)] Compute the overall $p$-value defined in equation (\ref{eq:test-Fisher}) and hence $Q$ defined in (\ref{eq:test-Pearson}).
\end{enumerate}
We use $B=10000$ in both randomization and permutation tests.
%which combines the $p$-values from each
%individual test to obtain an overall test statistics. Suppose we have carried out $J$ individual
%tests separately and resulted in $J$ $p$-values: $p_1,\dots,p_J$. Then t

The choice between Fisher's method and Pearson's method depend on the problem
at hand and \citet{owen2009} provides a review of both methods. In particular, 
if we suspect there are large
deviations from individual null hypotheses,
then Pearson's method has higher power than Fisher's method;  the paper 
also shows that the Pearson's method is admissible.
%is strong evidence that the directions of these individual hypotheses are almost the same,
%Both Fisher's and Pearson's methods are used in meta-analysis.
%\textcolor{blue}{The major difference between the two compare-then-combine approaches is that
%the second approach (meta-analysis methods) does not belong to the UIT family. In the UIT, to accpet
%$H_0$, we need all $p$-values to be greater than 0.05. However, for the Fisher's and Pearson's methods,
%we can have several $p_j < 0.05$, i.e. we reject $H_0$ in some cases, but still reject $H_0$ at
%the final test.}

The difference between the feature selection (see Section \ref{sec:test:uit}) and the 
Fisher's and Pearson's methods is that in the first approach,
we record the test statistics resulted in each single test and use an overall test statistics
to combine them together. Hence, obtaining a final $p$-value or carrying out feature selection
are both possible. In the second approach, we combine the $p$-values obtained from each individual
test together via a pre-defined combination function and calculate a final $p$-value. We cannot
proceed the feature selection by using this method.
The applications of the Fisher's and Pearson's methods on the two illustrative smile examples
are given in Section \ref{sec:compare-combine-Data1}, 
\ref{sec:compare-combine-Data2} and \ref{sec:compare-combine-Data1-2} respectively.

\section{Analyses of the Cleft Lip Smile Data}\label{sec:smile}
\subsection{Details}
\begin{figure}[!h]
    \centering
    \includegraphics[scale=.35,angle=0]{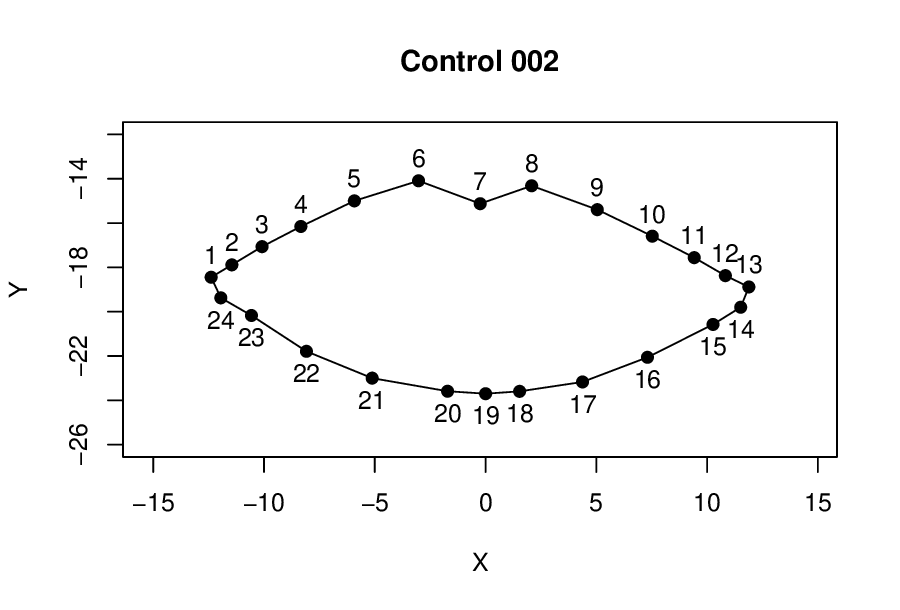}
    \captionsetup{width=1.0\textwidth}
    \caption{Landmark indices for $x$-$y$ coordinates on the lip periphery of a control subject at first frame from Smile Data 1.}
    \label{con002}
  \end{figure}

We now illustrate the methods using the cleft lip smile data. 
The data is collected using the Di4D system \citep{BV2023} as described in Section \ref{sec:asym} and it has been pre-registered using the basis registration.
This  data contains $N=25$ subjects, of which $N_1=12$ are control
subjects and $N_2=13$ are cleft subjects. There are $K=24$ landmarks
on the lip periphery. Figure \ref{con002} shows the lip
configuration in the $x$-$y$ plane. There are $K_P=11$ landmark pairs and $K_S=2$ solo
landmarks. The landmark indices are summarized in Table
\ref{lmindsmile}. The $M=3$ dimensional coordinates of these landmarks
have been extracted  at three frames: first (closed lip), middle (middle of
the smile) and last frame (maximum open lip smile). We have $J=MK_P+K_S=35$.
Let $g=1, 2$ denote the control and cleft groups, respectively.

\begin{table*}[t!]
\centering
\captionsetup{width=1.0\textwidth}
\caption{Indices of landmark pairs and solos in Figure \ref{con002} on the lip periphery for the Smile Data 1.}
\label{lmindsmile}
%\vspace{3mm}
\begin{tabular}{|c|>{\centering\arraybackslash}p{8cm}|} 
 \hline
 Landmark notation & Indices in Figure \ref{con002} \\
 \hline
  pair $(k_L,k_R) $ &  (1,13), (2,12), (3, 11), (4,10), (5,9), (6,8),\\
    &                (20,18), (21,17), (22,16), (23,15), (24,14)\\
 \hline
$k_S$ & 7, 19 \\
\hline
\end{tabular}
\end{table*}

  In the rest of this section several test statistics as described in Section \ref{sec:test}  will be presented under two headings   
  ``combine-then-compare approach" and ``compare-then-combine approach".
\subsection{Test Results for Smile Data 1}
\subsubsection{Application of the Combine-then-compare Approach}\label{sec:combine-compare-Data1}
 \citet{BV2023} carried out
  standard two-sided two-sample t-tests using the statistic
  (\ref{as1}), with conventional t-tables used to judge significance.
  Here we report the results of the one-sided
  versions of these t-tests since based on hypotheses in (\ref{originH0H1}), 
  one-sided tests are more appropriate. The results of our t-tests are given in Table
  \ref{meanvarasyBVpaper}.
  We reach the same conclusions as in \citet{BV2023} for their two-sided tests, namely, $H_0$ is
  rejected at all three frames. Further, the one-sided t-test indicates that
  there is more asymmetry for cleft lip subjects versus controls.
  Note that here the $p$-value is the smallest at the
  first frame, which suggests that the two groups are the most different
  at the beginning of the smile.

  \begin{table}[t!]
\begin{center}
\captionsetup{width=1.0\textwidth}
\caption{Mean, variance, t-values and $p$-values for $\phi^*_{L_1} (\bm a)$ using Smile Data 1 (equation (\ref{as1})) for cleft and control subjects at first, middle and last frames (\citet{BV2023}).}
\label{meanvarasyBVpaper}
%\vspace{3mm}
\begin{tabular}{c | c c c c c c c} 
 \hline
 \multicolumn{1}{c|}{} & \multicolumn{2}{|c|}{Cleft} & \multicolumn{2}{|c|}{Control} & \multicolumn{1}{c|}{} & \multicolumn{1}{|c}{One-sided test} & \multicolumn{1}{|c}{Two-sided test} \\
 \hline
  & mean & sd & mean & sd & $t$-values & $p$-values & $p$-values \\ [0.5ex] 
 \hline\hline
 First & 21.35 & 5.54 & 13.39  & 5.11  & -3.72  & 0.0006(***) & 0.0011(**) \\
 \hline
 Middle & 23.57 & 8.37 & 17.35  & 5.87  & -2.13 & 0.02(*) & 0.04(*) \\
 \hline
 Last & 25.65 & 9.66 & 18.22 & 5.21 & -2.36 & 0.01(*) & 0.03(*) \\
 \hline
\end{tabular}
\end{center}
\scriptsize{(*) = significant at the $5\%$ significance level. (**) = significant at the $1\%$ significance level. (***) = significant at the $0.1\%$ significance level.}
\end{table}

%choices $\psi_1 = a$ and $\psi_2 = a^2$, with the names$L_1$, $L_2$, respectively for the corresponding statistics.The weights are all 1 for both landmark pairs and solos.
%{\bf Combine-then-compare approach for Smile Data 1} 
Then, we use the
scores $\phi_{L_1}(\bm a)$ and $\phi_{L_2}(\bm a)$ defined in equations
(\ref{eq:L1score}) and (\ref{eq:L2score}) respectively.
Several one-sided two-sample Mann-Whitney U tests have been performed and the
results are shown in Table \ref{pwilxocon}. For the last two rows in the table, 
the weighted $\phi_{L_1}(\bm a)$ and weighted $\phi_{L_2}(\bm a)$ are used and $w_j$ are chosen
in an adaptive way as the following:
\begin{enumerate}
\item[(a)] Compute the sample mean shape on the doubled dataset, which
  contains basis registered original configurations and their reflections through
  midplane. Since the data has been pre-registered, the sample mean shape 
  is simply the arithmetic mean for each landmark. 
  %[XW: I am confused about whether
  %you are doing axis or basis registration here.  What happened to the
  %expert registration?]
  
    \item[(b)] The Euclidean distance within each landmark pair for 
    this mean shape is computed and its reciprocal is used as 
    the weight. The weights are the same for all 25 subjects.
    %among these subjects from both groups. 
    The weights for solo landmarks are the unit length based on the scale of data
    ($w_j=1$ in (\ref{eq:L1score}) and (\ref{eq:L2score})).
      %[XW: this choice depends on the units used to
      %measure distance.  A bit more care in the justification is needed.]
    \end{enumerate}
    %[XW:  You need to motivate to the
    %reader why t-test was used earlier and Mann-Whitney here. Also motivate
    %the choice of weights.]  

Different landmarks carry different amount of asymmetry information. 
Those landmarks near the midplane $\mathcal{P}$ may carry the most significant information 
on asymmetry according to the expert's knowledge. Thus, we give higher weights to landmarks 
near $\mathcal{P}$ and smaller weights on landmarks faraway from the central.

\begin{table*}[t!]
\begin{center}
\captionsetup{width=1.0\textwidth}
\caption{$p$-values from one-sided two-sample Mann-Whitney U tests using $\phi_{L_1}(\bm a)$ (\ref{eq:L1score}) and $\phi_{L_2}(\bm a)$ (\ref{eq:L2score}) based on Smile Data 1.}
\label{pwilxocon}
%\vspace{3mm}
\begin{tabular}{c | c c c}
\hline
 Composite asymmetry score & First frame & Middle frame & Last frame \\ [0.5ex] 
 \hline\hline
$\phi_{L_1}(\bm a)$ & 0.0001(***) & 0.043(*) & 0.011(*) \\
 \hline
$\phi_{L_2}(\bm a)$ & 0.0008(***) & 0.026(*) & 0.020(*) \\
 \hline
weighted $\phi_{L_1}(\bm a)$ & 0.0008(**) & 0.004(**) & 0.002(**) \\
 \hline
weighted $\phi_{L_2}(\bm a)$ & 0.002(*) & 0.015(*) & 0.006(**) \\
 \hline
\end{tabular}
\end{center}
\scriptsize{(*) = significant at the $5\%$ significance level. (**) = significant at the $1\%$ significance level. (***) = significant at the $0.1\%$ significance level.}
\end{table*}
%\footnotetext{\label{note1}The indication (*) means a result is significant at the $5\%$ significance level. The indication (**) means a result is significant at the $1\%$ significance level. The indication (***) means a result is significant at the $0.1\%$ significance level.}

It can be seen from Table \ref{pwilxocon} that $H_0$ is rejected in all cases.
In summary, there are significant differences between cleft and
control subjects at all frames, with control subjects less asymmetric
than cleft subjects overall. 
%The $L_1$ and $L_2$ statistics have a similar performance to one another. [XW: is this right?].

%[XW: write
%this paragraph in the style of earlier ones, with the SCIs appearing
%after the test results.]\\
\subsubsection{Application of the Compare-then-combine Approach}\label{sec:compare-combine-Data1}

{\bf Feature selection}
The method 
described in Section \ref{sec:test:uit} is used to select the landmarks 
which possess significant asymmetry information. 
$v_j$ is selected to be the two-sample t-statistics (equation (\ref{eq:v-stat})).
$V$ defined in (\ref{eq:max}) is computed and its critical value 
at $5\%$ significance level is determined via bootstrap. 
The total number of bootstrap iterations used was 10000. \citet{boyett1977} has suggested
17000 iterations to make sure the 
significance level is $1\%$. However, the results of using 
10000 and 17000 iterations turned out to be the same for our Smile Data 1. 
The central pair, landmarks 6 and 8, corresponds to $t$-values 
which exceed these critical points at all three frames.
Thus, these two landmarks carry the
most significant information of asymmetry. 
%Further, the simultaneous confidence
%interval related to landmarks 6 and 8 excludes 0, so that corresponding $H_{0j}$ is rejected
%for these landmarks.

\begin{figure}[h!]
        \centering
        \includegraphics[scale = 0.3]{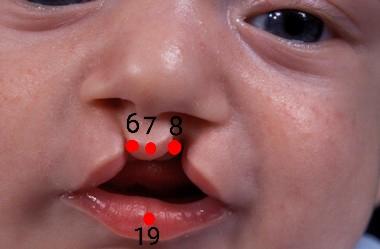}
        %\vspace{3mm}
        \captionsetup{width=1.0\textwidth}
        \caption{Photo of a cleft patient with central pair landmarks and solo landmarks. This figure is reproduced from photo taken from the website {\url{https://www.nhs.uk/conditions/cleft-lip-and-palate/}}.}
        \label{cleftpatient}
    \end{figure}

Further, most of the extreme $p$-values obtained from Mann-Whitney U tests 
are related to landmarks 6, 8 and two solo landmarks 7 and 19. These tests are performed
using $\phi_{L_1}$ and $\phi_{L2}$ computed on various subsets of landmarks.
This suggests that the two groups are differed at most when using these four landmarks. So these
landmarks can carry important information on asymmetries. This
finding matches the expert's opinion. Figure \ref{cleftpatient} shows
a photo of a cleft patient, where the central pair (6, 8) and solo
landmarks, 7 and 19, are marked on the photo and the landmarks are
same as in Figure \ref{con002}. Roughly, it is a `Y'-shape if we join
the landmarks 6, 7, 8 and then 7 with 19 by line segments. From medical point of view, this
particular shape is known a priori to be the most important for the asymmetry assessment.
%by the orthodontic surgeons.

%\textbf{Done up to here}

\textbf{Meta-analysis}
Fisher's method and Pearson's method defined in equation (\ref{eq:test-Fisher})
and (\ref{eq:test-Pearson}) respectively are performed. Mann-Whitney U test is used to
obtain individual left-sided $p$-values, $\Tilde{p}_j$, for $j=1,\dots,35$,
using $\bm a_n$, $n=1,\dots,25$.

For Fisher's method, we compare $P_L$ with the tail of $\chi_{70}^2$ distribution (as $J=35$)
to obtain overall $p$-values.
For the Pearson's method, randomization tests with 10000 iterations are used to
estimate the overall $p$-values. The values of $P_L$, $P_R$, $Q$ and $p$-values derived
at the three frames are shown in Table \ref{tab:meta-result}. $H_0$ is rejected
at all frames when we use Fisher's method. For Pearson's method, $H_0$ is rejected at
the first and last frames. Comparing with Table \ref{meanvarasyBVpaper} and
\ref{pwilxocon}, the $p$-values for the Fisher's method are more extreme. This indicates
that Fisher's method is more sensitive to the departures of cleft group from control group.

\begin{table*}[t!]
    \begin{center}
        \captionsetup{width=1.0\textwidth}
    \caption{The results of $P_R$ and $p$-values obtained from Fisher's method and Pearson's method at all three frames using the Smile Data 1.}
    \label{tab:meta-result}
    %\vspace{2mm}
    \begin{tabular}{c|c c c c c }
    \hline
         & $P_L$ & $P_R$ & $Q$ & Fisher's method & Pearson's method \\
         \hline\hline
        First frame & 206.65 & 12.26 & 206.65 & 0(***) & $4 \times 10^{-4}$(***) \\
         \hline
        Middle frame & 127.19 & 33.90 & 127.19 & $3.51 \times 10^{-5}$(***) & 0.067 \\
         \hline
        Last frame & 134.21 & 26.89 & 134.21 & $6.18 \times 10^{-6}$(***) & 0.027(*) \\
         \hline
    \end{tabular}
    \end{center}
\scriptsize{(*) = significant at the $5\%$ significance level. (**) = significant at the $1\%$ significance level. (***) = significant at the $0.1\%$ significance level.}
\end{table*}

\subsubsection{Welch's t-test and Hotelling's $T^2$ Test}
%\textbf{\textcolor{red}{As a subsection or a heading?}}
There are a few other tests we have applied and could be relevant for some future applications
\begin{itemize}
    \item[(a)] The Welch version of the $t$-test (\citet{welch1947}, \citet{lehmann2022}),  which accommodates different variances for the two groups. In this example, the results of
the Welch $t$-test is found to be similar to the standard $t$-test 
(Table \ref{meanvarasyBVpaper}). A possible reason is that our sample size is small, 
so it results in similar outcomes.
    %Welch's t-test,
    \item[(b)] Hotelling's $T^2$ test \citep{kvm2024}. We apply here Hotelling's $T^2$ test on unsigned elementary 
feature vector $\bm a_n$, $n=1,\ldots,N$. In this example, we reject $H_0$ for the first two frames at $5\%$ significance level.
%fail to reject $H_0$ at the last frame the test results fail to improve the outcomes of the 
%u%nivariate tests.
The reason could be that a Hotelling's $T^2$
test treats the features separately so it  does not gain any power if
the features point in the same direction; one-sided Hotelling's $T^2$ could be an appropriate approach, which we will pursue in future.
In this example of Smile Data 1, the sample size is less than the dimension of $\bm a_n$, 
i.e. $N<J$, so the covariance matrix is singular, which requires some modifications and will be dealt in future.
%\textcolor{red}{for example, most of the components of $\Bar{\bm a}^{(1)}-\Bar{\bm a}^{(2)}$ point towards negative part of the axis, where $\Bar{\bm a}^{(g)}$ is the mean vector of $\bm a_n$ for group $g=1,2$}
\end{itemize}
%We also applied  
 %Further, the multivariate Hotelling's $T^2$ test  is an alternative approach 
%to the UIT. 

\subsection{Visual Study Related to Three Composite Statistics}\label{subsec:dotplot}
%\textbf{\textcolor{red}{Dot plots can be moved after the feature selection.}}

 \begin{figure}[!h]
 \center
  \begin{subfigure}[b]{0.3\textwidth}
         \centering
         \includegraphics[width=\textwidth]{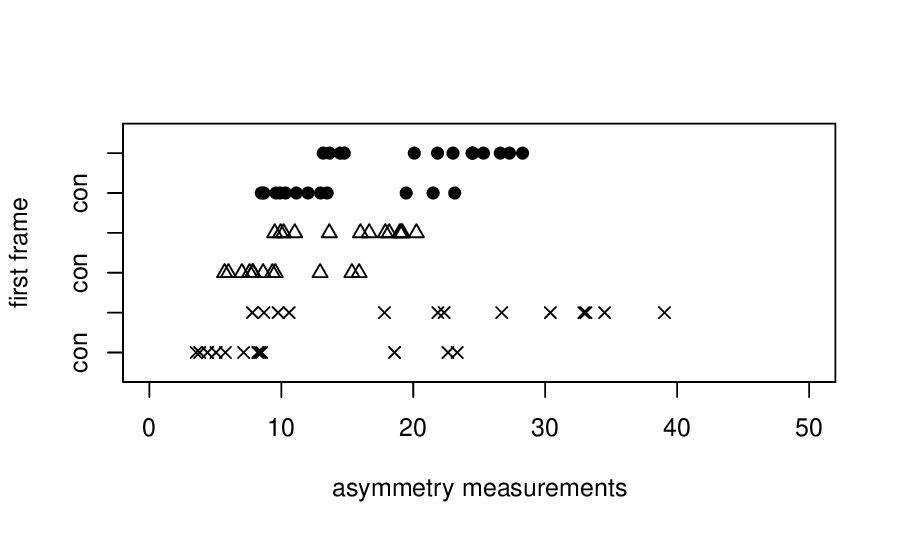}
         \subcaption{First frame}
     \end{subfigure}
     \begin{subfigure}[b]{0.3\textwidth}
         \centering
         \includegraphics[width=\textwidth]{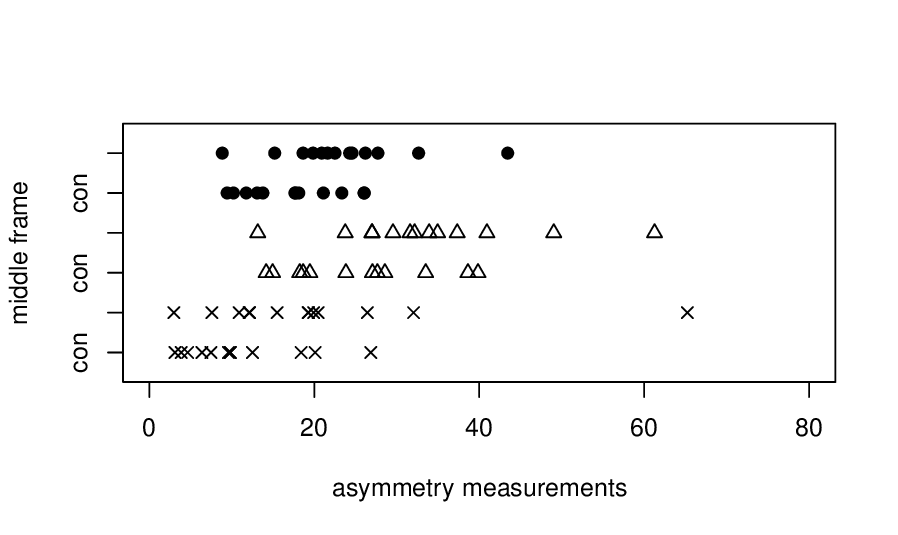}
         \subcaption{Middle frame}
     \end{subfigure}
          \begin{subfigure}[b]{0.3\textwidth}
         \centering
         \includegraphics[width=\textwidth]{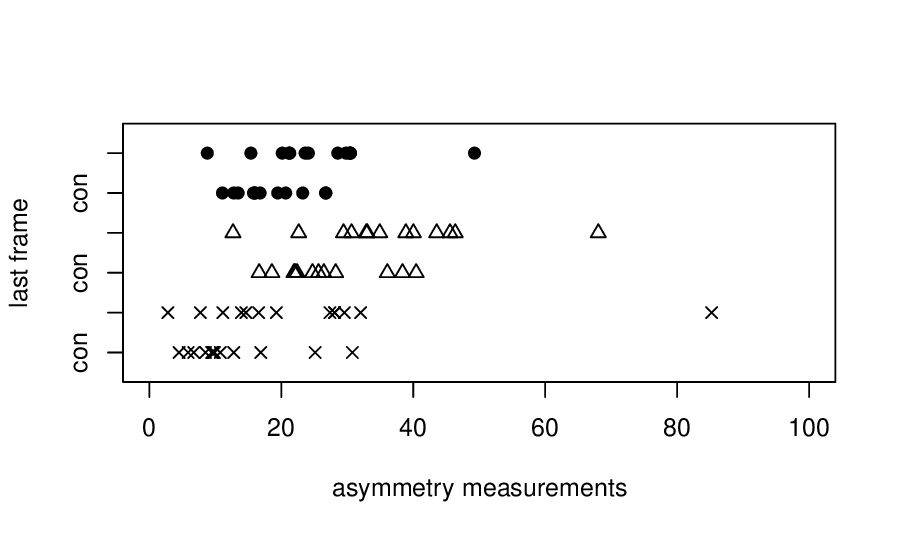}
         \subcaption{Last frame}
     \end{subfigure}
     %\vspace{3mm}
     \captionsetup{width=1.0\textwidth}
  \caption{Plots for three composite asymmetry scores. From top to bottom, we have $\phi^*_{L_1} (\bm a)$ (\ref{as1}) (in solid dots), $\phi_{L_1}(\bm a)$ (\ref{eq:L1score}) (in triangles) and $\phi_{L_2}(\bm a)$ (\ref{eq:L2score}) (in crosses) at the three frames computed on the Smile Data 1. Plot (a) is at first frame, while (b) is at middle frame, whereas (c) is at last frame. In plot (a), $\phi_{L_1}(\bm a)$ and $\phi_{L_2}(\bm a)$ are divided by 2. In plot (b) and (c), $\phi_{L_2}(\bm a)$ is divided by 3. The first, third and fifth rows from top to bottom are cleft while others are control. }
  \label{AS1L2L2dot}	     
  \end{figure} 

Since the test results indicate significant differences between the two groups,
we would like to visualize the distributions of $\phi^*_{L_1}(\bm a)$, 
$\phi_{L_1}(\bm a)$ and $\phi_{L_2}(\bm a)$ defined in (\ref{as1}),
(\ref{eq:L1score}) and (\ref{eq:L2score}) respectively between the two groups. 
Hence, dot plots similar to the stem 
leaf plot are created and given in Figure \ref{AS1L2L2dot}.
In these dot plots, composite asymmetry 
scores for both groups are plotted together at each frame.
According to these dot plots, substantial overlaps between cleft and control 
subjects can be seen. Moreover, the three scores used in this 
section all have similar performances, as the separation between the two groups 
at each frame revealed by the dot plots are all similar. 
%Figure \ref{asydot} shows the dot plots for  
%defined in equation , whereas Figure \ref{L1dotplot} and 
%\ref{L2dotplot} display the dot plots of $\phi_{L_1}(\bm a)$ 
%and $\phi_{L_2}(\bm a)$ (equations ). 

\begin{figure}[!h]
 \center
  \begin{subfigure}[b]{0.3\textwidth}
         \centering
         \includegraphics[width=\textwidth]{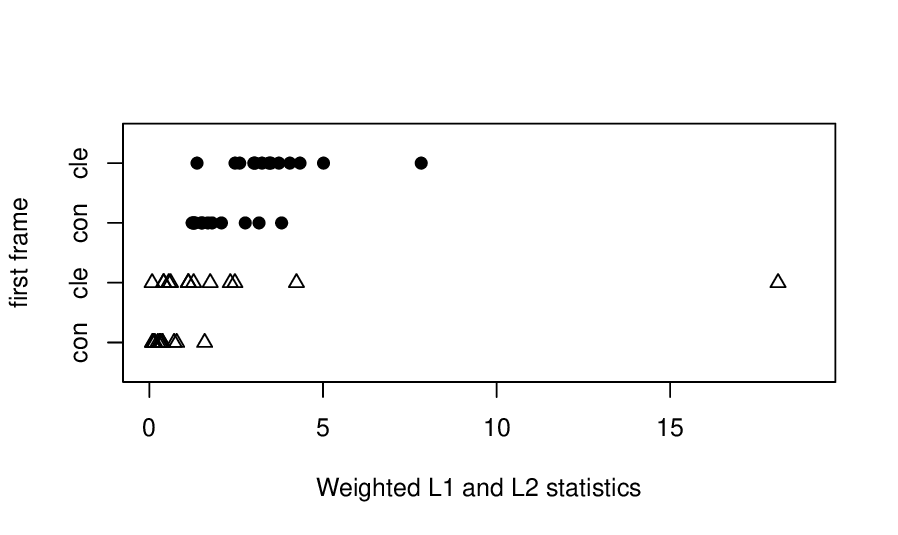}
         \subcaption{First frame}
     \end{subfigure}
     \begin{subfigure}[b]{0.3\textwidth}
         \centering
         \includegraphics[width=\textwidth]{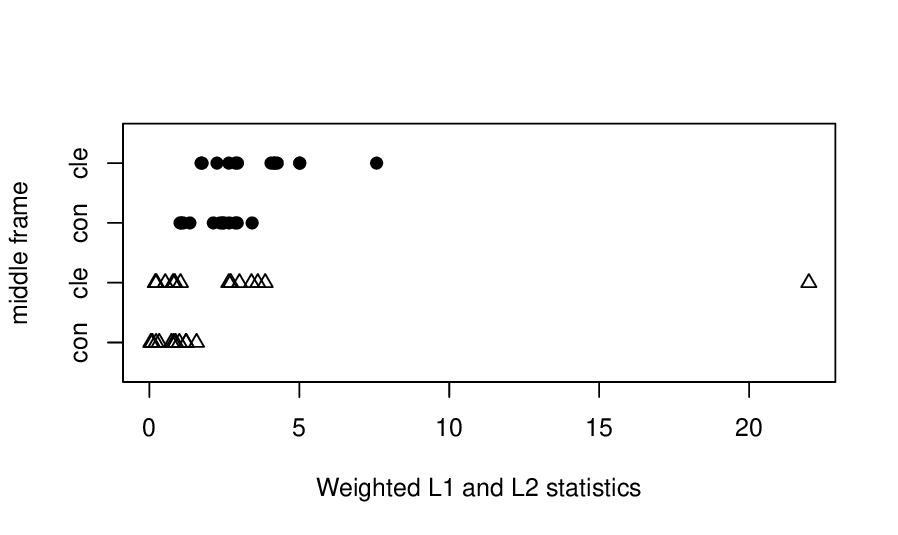}
         \subcaption{Middle frame}
     \end{subfigure}
          \begin{subfigure}[b]{0.3\textwidth}
         \centering
         \includegraphics[width=\textwidth]{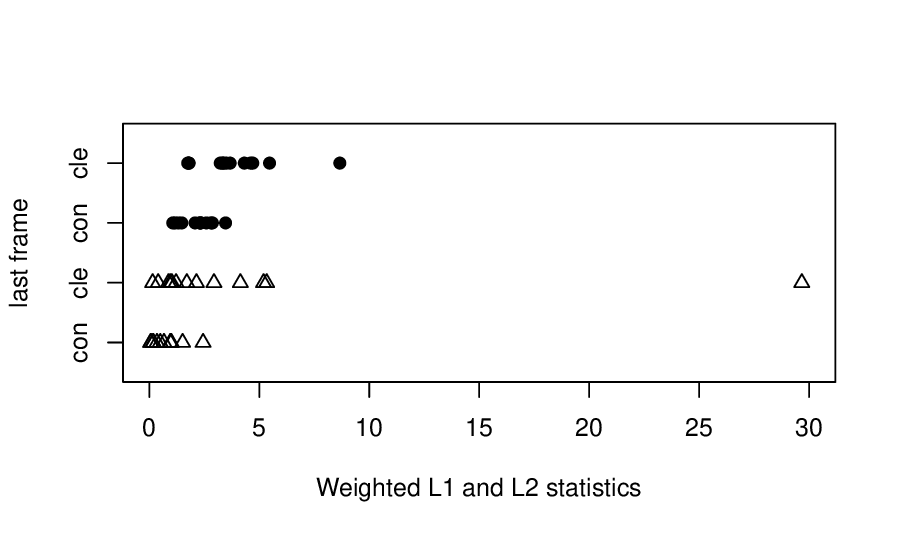}
         \subcaption{Last frame}
     \end{subfigure}
     %\vspace{3mm}
     \captionsetup{width=1.0\textwidth}
  \caption{Plots of two weighted asymmetry scores. From top to bottom, we have weighted $\phi_{L_1}(\bm a)$ (in solid dots) and $\phi_{L_2}(\bm a)$ (in triangles) at the three frames computed using the Smile Data 1. The first and third rows from top to bottom are cleft while others are control. Plot (a) is at first frame, while (b) is at middle frame, whereas (c) is at last frame.}
  \label{weightL1dotplot}	     
  \end{figure}

Figure \ref{weightL1dotplot} shows the dot plots for weighted $\phi_{L_1}(\bm a)$
and weighted $\phi_{L_2}(\bm a)$ simultaneously. According
to these figures, these two composite scores 
push outliers farther away from the rest of the samples, more so for weighted $\phi_{L_1}$.
%and $\phi_{L_2}$. 
%\textcolor{red}{So the weighted $\phi_{L_1}(\bm a)$ 
%and weighted $\phi_{L_2}(\bm a)$ emphasize which are more extreme in cleft lip group. %(\textbf{redundant})}

 % \textbf{For Figure 4 and 5, repeat what are the (a), (b) and (c). Also start with saying "Plots for 3 composite statistics"}

%The top row of the graph is for cleft lip subjects whereas the bottom row is for control subjects.
%     \begin{figure}[!t]
% \center
%  \begin{subfigure}[b]{0.3\textwidth}
%        \centering
 %        \includegraphics[width=\textwidth]{L2weight_fir.eps}
 %        \subcaption{First frame}
  %   \end{subfigure}
   %  \begin{subfigure}[b]{0.3\textwidth}
   %      \centering
   %      \includegraphics[width=\textwidth]{L2weight_mid.eps}
   %      \subcaption{Middle frame}
   %  \end{subfigure}
   %       \begin{subfigure}[b]{0.3\textwidth}
   %      \centering
   %      \includegraphics[width=\textwidth]{L2weight_la.eps}
   %      \subcaption{Last frame}
   %  \end{subfigure}
   %  \vspace{3mm}
   %  \captionsetup{width=1.0\textwidth}
  %\caption{Dot plots of weighted $\phi_{L_2}(\bm a)$ at the three frames. The top row of the graph is for cleft lip subjects whereas the bottom row is for control subjects.}
  %\label{weightL2dotplot}	     
  %\end{figure} 
%This indicates that some surgeries are successful and one should expect
%this. Nevertheless, some cleft subjects are still abnormal.
%[XW: I think it is dangerous to impute surgical reasons here.  Just report
%the facts.]

%{\bf This section is still open to editing; add wherelse we can use our method? Procrustes? Future direction?}
%\textbf{\textcolor{red}{A subsection a heading for summary?}}]
\subsection{Summary of the Analyses}
We end this section by summarizing the main conclusions for these cleft lip data:
\begin{itemize}
    \item[(a)] There are statistically significant differences between the two groups at all frames. 
    %The control subjects are overall less asymmetric than the cleft lip subjects.
    The control subjects are overall more symmetric than the cleft lip subjects.
    \item[(b)] Substantial overlaps can be found on dot plots in Figure \ref{AS1L2L2dot}. This matches the expert's expectation that some cleft lip subjects should be fairly close to normal subjects after surgeries.
    \item[(c)] The landmarks 6, 7, 8 and 19 (the `Y'-shape) are the most important in assessing asymmetries.
\end{itemize}

\section{Analyses of the Orthognathic Surgery Smile Data}\label{sec:jaw}
\subsection{Details}
\begin{figure}[!ht]
    \centering
      \begin{subfigure}[h]{0.4\textwidth}
         \centering
         \includegraphics[scale = 0.3]{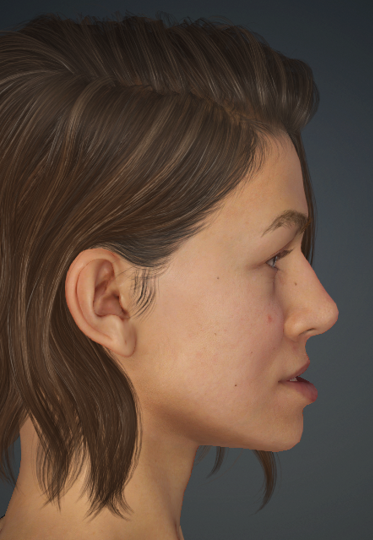}
     \end{subfigure}
     \hspace{-6mm}
     \begin{subfigure}[h]{0.4\textwidth}
         \centering
         \includegraphics[scale = 0.3]{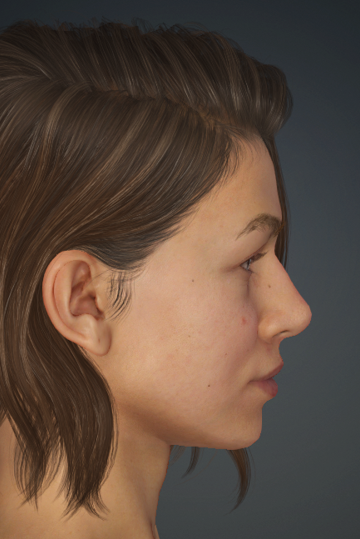}
     \end{subfigure}
     \vspace{3mm}
     \captionsetup{width=1.0\textwidth}
  \caption{View of the face from right before (left) and after (right) the orthognathic surgery.}
  \label{fig:face}
  \end{figure}

%There is another dataset, the orthognathic data, considered by us, 
%which is referred to as the jaw data. It is collected
%and pre-registered in the same way as the cleft lip data. 
The orthognathic surgery smile data contains $N=22$ subjects who have
undergone orthognathic surgery and their pre- and post-surgery data have both been collected. 
Figure \ref{fig:face} shows the face before (in the left) and after (in the right) the
orthognathic surgery, where the view is from the right side of the face. It can be seen in the left of Figure \ref{fig:face}, the jaw is somewhat protruded in comparison to the figure on right.

%\textbf{What does one mean by orthognathic data and what surgery is going to achieve? To get the Figure which we have from BSK of the subject before and after and to put it in the Supplementary at least.}
%The correction of the 
%anterior-posterior skeletal and soft tissue positions should not result in detrimental changes 
%in the transverse dimension, i.e. asymmetries of the dynamic movement of the nasolabial region.

\begin{figure}[!ht]
    \centering
      \begin{subfigure}[t]{0.2\textwidth}
         \centering
         \includegraphics[scale = 0.33]{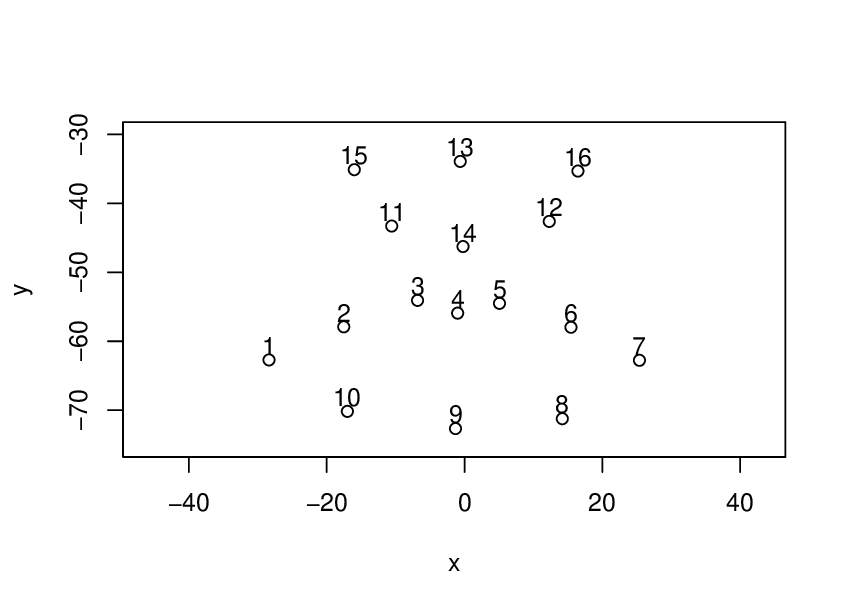}
     \end{subfigure}
     \hspace{16mm}
     %\vfill
     \begin{subfigure}[t]{0.2\textwidth}
         %\centering
         \includegraphics[scale = 0.35]{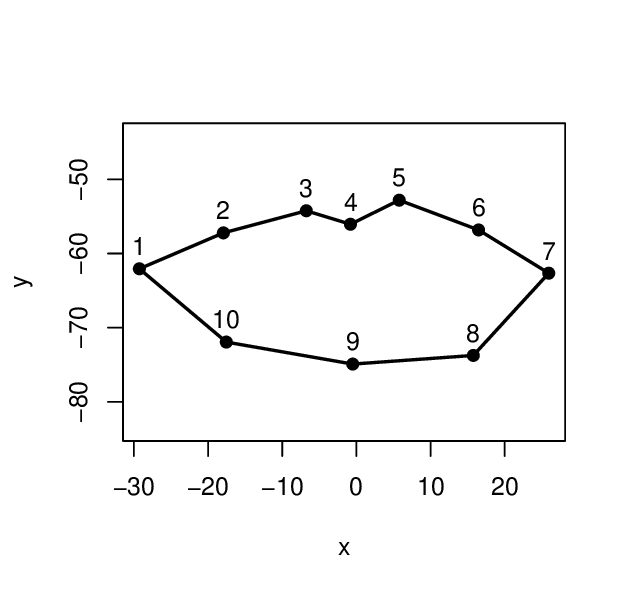}
     \end{subfigure}
     %\vspace{3mm}
     \captionsetup{width=1.0\textwidth}
  \caption{The left figure shows the 16 landmarks on the nasolabial region for a typical subject at the first frame using the pre-surgery data. The right figure shows the 10 landmarks on the lip periphery. The two figures are based on the Smile Data 2.}
  \label{fig:lmjaw}
  \end{figure}

The axes used for the Smile Data 2 are exactly the same as the Smile Data 1 and the details are
given in Section \ref{sec:asym:reg}. The 16 landmarks on the nasal region are shown in left of Figure
\ref{fig:lmjaw}, while the right of Figure \ref{fig:lmjaw} shows the 10 landmarks on 
the lip. The landmark indices are summarized in Table \ref{tab:lmindjaw}. The scale of the
Smile Data 2 is in mm. The same frames as the Smile Data 1 are used.
%Note that the landmarks with indices 1 to 10 in Table \ref{tab:lmindjaw}
%pertains to the right plot of Figure \ref{fig:lmjaw}, whereas the landmarks with indices 15 to 20
%correspond to the left of Figure \ref{fig:lmjaw}.
%which are the same as the smile data

\begin{table*}[t!]
\centering
\captionsetup{width=1.0\textwidth}
\caption{Indices of landmark pairs and solos in Figure \ref{fig:lmjaw} on the lip periphery and nasolabial region for the Smile Data 2.}
\label{tab:lmindjaw}
%\vspace{3mm}
\begin{tabular}{|c|>{\centering\arraybackslash}p{8cm}|} 
 \hline
 Landmark notation & Indices in Figure \ref{fig:lmjaw} \\
 \hline
  pair $(k_L,k_R) $ &  (1,7), (2,6), (3, 5), (8,10), (11,12), (15,16)\\
 \hline
$k_S$ & 4, 9, 13, 14 \\
\hline
\end{tabular}
\end{table*}

There are two sets of landmarks we will focus on:
\begin{enumerate}
    \item[(a)] The 7 landmarks on the upper lip (landmarks 1 to 7 in Figure \ref{fig:lmjaw}) and the 6 landmarks on the nasal region (landmarks 11 to 16 in the left of Figure \ref{fig:lmjaw}). There are $K_P=5$ paired landmarks and $K_S=3$ solo landmarks, hence, $J=18$.
    \item[(b)] The 10 landmarks on the lip (landmarks 1 to 10 in Figure \ref{fig:lmjaw}). There are $K_P=4$ paired landmarks and $K_S=2$ solo landmarks, hence, $J=14$.
\end{enumerate}

These two sets of landmarks correspond to two aims:
\begin{enumerate}
    \item[(a)] Assess the affects of surgery, by comparing pre- and post-surgery data via paired tests (item (a) above).
    \item[(b)] Compare the subjects from Smile Data 2 to the control subjects from the Smile Data 1 where we use the same landmarks on the lip for both data (item (b) above).
\end{enumerate}
Some previous studies have been
carried out, such as in \citet{alhiyali2015}, \citet{xue2020}, \citet{xue2023} and 
\citet{quast2023}. These studies have focused on different facial expressions and the authors have taken
average over measurements resulted from these expressions. However, by doing this, they would
lose a lot of information. Our work is different from these previous studies as we 
attempt to construct asymmetry measurements over the lip and nasal region at different
frames during the smile.

The rest of this section will be divided into two parts corresponding to
item (a) (effect of surgery, see Section \ref{sec:surgeryeffect}) and 
item (b) (deviation from normal subjects, see Section \ref{sec:deviationcontrol}) above.

\subsection{Test Results for Smile Data 2: Effect of Surgery}\label{sec:surgeryeffect}
The pre-surgery data is compared to the post-surgery data by using one-sided paired tests on
the asymmetry measurements. Let $g=1, 2$ denote the pre-surgery data
and post-surgery data respectively. The absolute elementary feature vectors $\bm a_n^{(1)}$ and
$\bm a_n^{(2)}$ are computed using the pre- and post-surgery data separately, for $n=1,\dots,22$.
%Since $K_P=5$ and $K_S=3$, we have $J=18$.

%\textbf{}
\subsubsection{Application of the Combine-then-compare Approach}\label{sec:combine-compare-Data2}
The 13 landmarks on the upper lip and nasal region
are used. Paired t-tests have been performed on composite
asymmetry scores defined in equation (\ref{as1}) whereas Wilcoxon 
signed-rank tests have been performed on scores defined in (\ref{eq:L1score}), (\ref{eq:L2score})
respectively.

Since the expert suspects that the asymmetries are increased after the surgery, so the one-sided
alternative hypothesis becomes the distribution relates to pre-surgery data is stochastically
smaller than that of post-surgery data. The Group 1 and Group 2 in (\ref{originH0H1}) are changed
to pre- and post-surgery data respectively.

The test results on $\phi^*_{L_1}$ given in equation (\ref{as1}) are shown in Table 
\ref{tab:as1result}. $H_0$ is rejected in middle and last frames. In other words, the
asymmetries are increased after surgery when inspect at these two frames. Further,
the $p$-value is the most extreme at the last frame, which indicates that the two groups
are the most different at the last frame.
\begin{table*}[t!]
\begin{center}
\captionsetup{width=1.0\textwidth}
\caption{Mean, variance, t-values and $p$-values using $\phi^*_{L_1} (\bm a)$ based on the Smile Data 2 (equation (\ref{as1})).}
\label{tab:as1result}
%\vspace{3mm}
\begin{tabular}{c | c c c c c c} 
 \hline
 \multicolumn{1}{c|}{} & \multicolumn{2}{|c|}{Pre} & \multicolumn{2}{|c|}{Post} & \multicolumn{1}{c|}{} & \multicolumn{1}{|c}{One-sided paired test} \\
 \hline
 $\phi^*_{L_1} (\bm a)$ & mean & sd & mean & sd & $t$-values & $p$-values \\ [0.5ex] 
 \hline\hline
 First & 15.69 & 5.64 & 18.47 & 8.49 & -1.33  & 0.098 \\
 \hline
 Middle & 17.53 & 5.92 & 22.25 & 10.89 & -2.36 & 0.014(*) \\
 \hline
 Last & 17.72 & 5.45 & 22.77 & 9.52 & -3.29 & 0.002(**) \\
 \hline
\end{tabular}
\end{center}
\scriptsize{(*) = significant at the $5\%$ significance level. (**) = significant at the $1\%$ significance level. (***) = significant at the $0.1\%$ significance level.}
\end{table*}

Table \ref{tab:pwilxoconnewlm} displays the results of Wilcoxon signed-rank tests
on $\phi_{L_1}(\bm a)$ given by (\ref{eq:L1score}) and $\phi_{L_2}(\bm a)$ given by (\ref{eq:L2score}).
We reach the same outcomes as the paired t-tests: $H_0$ is rejected at the middle
and last frames and the two groups are the most different at the last frame.
%Further, the $p$-values are the smallest at the last frame.
\begin{table*}[t!]
\begin{center}
\captionsetup{width=1.0\textwidth}
\caption{$p$-values from one-sided Wilcoxon signed-rank tests using $\phi_{L_1}(\bm a)$ (\ref{eq:L1score}) and $\phi_{L_2}(\bm a)$ (\ref{eq:L2score}) based on the Smile Data 2.}
\label{tab:pwilxoconnewlm}
%\vspace{3mm}
\begin{tabular}{c | c c c}
\hline
 Composite asymmetry score & First frame & Middle frame & Last frame \\ [0.5ex] 
 \hline\hline
$\phi_{L_1}(\bm a)$ & 0.13 & 0.016(*) & 0.002(**) \\
 \hline
$\phi_{L_2}(\bm a)$ & 0.088 & 0.007(**) & 0.001(**) \\
 \hline
\end{tabular}
\end{center}
\scriptsize{(*) = significant at the $5\%$ significance level. (**) = significant at the $1\%$ significance level. (***) = significant at the $0.1\%$ significance level.}
\end{table*}

\subsubsection{Application of the Compare-then-combine Approach}\label{sec:compare-combine-Data2}
%\textbf{ approach for Smile Data 2}
\textbf{Feature selection}
The procedures given in Section \ref{sec:test:uit} are performed using pre- and
post-surgery data, in order to select which landmarks contribute the most to the increment
in asymmetries after the surgery.
10000 bootstrap iterations are performed. The $25\%$, $20\%$, $15\%$ and $10\%$
critical values of $V$ are determined at the three frames separately. The corner pair, landmarks (1, 7) are important
at the middle frame with respect to $25\%$, $20\%$ and $15\%$ critical values of $V$. The significance levels
used here are larger than we used for Smile Data 1 in Section \ref{sec:smile}, which implies
the features selected here are less important than the features in Smile Data 1.
%Table  shows the landmark indices which contribute the most to the asymmetries.

\textbf{Meta-analysis}
Fisher's method and Pearson's method given in (\ref{eq:test-Fisher})
and (\ref{eq:test-Pearson}) are performed with Wilcoxon signed-rank tests
as individual tests using $\bm a_n^{(1)}$ and $\bm a_n^{(2)}$, for $n=1,\dots,22$.
$P_L$ defined in equation (\ref{eq:test-Fisher}) is computed and compare with the tail
of $\chi^2_{36}$ distribution (as $J=18$). As for $Q$, permutation tests are used to estimate
the overall $p$-values. Table \ref{tab:pairedmeta-result} shows the values of 
$P_L$, $P_R$, $Q$ and $p$-values for Fisher's and Pearson's method at the three frames.

\begin{table*}[t!]
    \begin{center}
        \captionsetup{width=1.0\textwidth}
    \caption{The results of $P_R$ and $p$-values obtained from Fisher's method and Pearson's method at all three frames. Pre- and post-surgery data from the Smile Data 2 is used here.}
    \label{tab:pairedmeta-result}
    %\vspace{2mm}
    \begin{tabular}{c|c c c c c }
    \hline
         & $P_L$ & $P_R$ & $Q$ & Fisher's method & Pearson's method \\
         \hline\hline
        First frame & 45.52 & 20.04 & 45.52 & 0.13 & 0.45 \\
         \hline
        Middle frame & 63.39 & 10.09 & 63.39 & 0.003(**) & 0.085 \\
         \hline
        Last frame & 64.20 & 11.53 & 64.20 & 0.003(**) & 0.041(*) \\
         \hline
    \end{tabular}
    \end{center}
\scriptsize{(*) = significant at the $5\%$ significance level. (**) = significant at the $1\%$ significance level. (***) = significant at the $0.1\%$ significance level.}
\end{table*}

For the Fisher's method, we obtain the same results as in the combine-then-compare
approach: $H_0$ is rejected at middle and last frames. However, the $p$-values
obtained in Fisher's method are more extreme at the middle frame than those from combine-then-compare
approach. For the Pearson's
method, we fail to reject $H_0$ at the first and middle frames. This indicates that
the Pearson's method may not be sensitive to changes in asymmetries before
and after the surgery.

%\textbf{Compare-then-combine approach using (\ref{eq:test-Fisher}) and (\ref{eq:test-Pearson}):}
\subsection{Test Results for Smile Data 1 and Smile Data 2: Deviation from Control Subjects}\label{sec:deviationcontrol}
The Smile Data 2 is compared with the control subjects from the Smile Data 1.
In Section \ref{sec:smile}, we only use 12 subjects which form a subset of the
control group. The expert believes that more data should be used, thus, all 109
control subjects are used in this section.
To make the comparisons available, a subset of 10 landmarks on the lip (Figure \ref{fig:lmjaw}) are chosen from the
24 landmarks (see Figure \ref{con002}), so that the Smile Data 2 and the Smile Data 1 
have the same landmarks at the lip. Hence, the absolute elementary feature vectors $\bm a_n$
for $n=1,\dots,12$ are
recomputed on the subjects from Smile Data 1 using the new landmarks.
%One-sided two-sample
%t-tests have been performed on scores defined in (\ref{as1}) while Mann-Whitney U tests
%have been carried out on scores .
Pre- and post-surgery data are used
separately. Hence, we have three groups in total, but we stick with two-sample tests. 
We scale the data so that both datasets have the same scale in cm.

%\textbf{ }
%The asymmetry measurements are computed and corresponding two-sample tests are performed.
\subsubsection{Application of the Combine-then-compare Approach}\label{sec:combine-compare-Data1-2}
The one-sided two-sample t-tests are performed on $\phi^*_{L_1}$ defined in (\ref{as1}) and the outcomes are shown in
Table \ref{tab:as1resultjawsmilepre}. Table \ref{tab:as1meanvarjawsmile} shows the means and
variances of $\phi^*_{L_1}$ for the three groups of data. $H_0$ is rejected in almost all cases,
except for the pre-surgery data at middle frame. This implies that in most of the cases,
control subjects from the Smile Data 1
%(\textbf{\textcolor{red}{may need a name for the jaw data subjects}})
are more symmetric than subjects from the Smile Data 2 .
\begin{table}[!t]
\begin{center}
\captionsetup{width=1.0\textwidth}
\caption{$p$-values from one-sided two-sample Welch's $t$-test using $\phi_{L_1}^*$ defined in (\ref{as1}) to compare between Smile Data 2 subjects with normal subjects from the Smile Data 1.}
\label{tab:as1resultjawsmilepre}
%\vspace{3mm}
\begin{tabular}{c | c c c}
\hline
$\phi_{L_1}^*$  & First frame & Middle frame & Last frame \\ [0.5ex] 
 \hline\hline
Pre-surgery & 0.006(**) & 0.073 & 0.031(*) \\
 \hline
Post-surgery & 0.025(*) & 0.013(*) & 0.010(*) \\
 \hline
\end{tabular}
\end{center}
\scriptsize{(*) = significant at the $5\%$ significance level. (**) = significant at the $1\%$ significance level. (***) = significant at the $0.1\%$ significance level.}
\end{table}

\begin{table}[t!]
\begin{center}
\captionsetup{width=1.0\textwidth}
\caption{Mean, standard deviation, t-values of $\phi^*_{L_1} (\bm a)$ (equation (\ref{as1})) for pre-, post-surgery subjects from Smile Data 2 and 12 control subjects from Smile Data 1.}
\label{tab:as1meanvarjawsmile}
\vspace{2mm}
%\vspace{3mm}
\begin{tabular}{c | c c c c c c} 
 \hline
 \multicolumn{1}{c|}{} & \multicolumn{2}{|c|}{Pre} & \multicolumn{2}{|c|}{Post} & \multicolumn{2}{c}{Control} \\
 \hline
 $\phi^*_{L_1} (\bm a)$ & mean & sd & mean & sd & mean & sd \\ [0.5ex] 
 \hline\hline
 First & 0.95 & 0.35 & 0.87 & 0.29 & 0.74  & 0.22 \\
 \hline
 Middle & 1.17 & 0.41 & 1.30 & 0.52 & 1.03 & 0.35 \\
 \hline
 Last & 1.22 & 0.34 & 1.32 & 0.46 & 1.07 & 0.33 \\
 \hline
\end{tabular}
\end{center}
\end{table}

\nil{\begin{table}[t!]
\begin{center}
\captionsetup{width=1.0\textwidth}
\caption{Mean, variance, t-values of $\phi^*_{L_1} (\bm a)$ (equation (\ref{as1})) for pre-, post-surgery subjects from Smile Data 2 and control subjects from Smile Data 1.}
\label{tab:as1meanvarjawsmile}
%\vspace{3mm}
\begin{tabular}{c | c c c c c c} 
 \hline
 \multicolumn{1}{c|}{} & \multicolumn{2}{|c|}{Pre} & \multicolumn{2}{|c|}{Post} & \multicolumn{2}{c}{Control} \\
 \hline
 $\phi^*_{L_1} (\bm a)$ & mean & sd & mean & sd & mean & sd \\ [0.5ex] 
 \hline\hline
 First & 0.32 & 0.12 & 0.29 & 0.10 & 0.25  & 0.07 \\
 \hline
 Middle & 0.39 & 0.14 & 0.43 & 0.17 & 0.34 & 0.12 \\
 \hline
 Last & 0.41 & 0.11 & 0.44 & 0.15 & 0.36 & 0.11 \\
 \hline
\end{tabular}
\end{center}
\end{table}}

One-sided two-sample Mann-Whitney U tests have been performed using $\phi_{L_1}$
and $\phi_{L_2}$ defined in (\ref{eq:L1score}) and (\ref{eq:L2score}) respectively.
The test results are shown in Table \ref{tab:pwilcoxon109con}.
$H_0$ is rejected in almost all cases except for pre-surgery data at the middle frame and post-surgery data at the first frame.
%The outcomes are the same as the t-tests: $H_0$ is rejected in all cases except
\begin{table}[!t]
\begin{center}
\captionsetup{width=1.0\textwidth}
\caption{$p$-values obtained from one-sided two-sample Mann-Whitney U tests using $\phi_{L_1}$ in (equation (\ref{eq:L1score})) and $\phi_{L_2}$ in (\ref{eq:L2score}) to compare between Smile Data 2 subjects with control subjects from the Smile Data 1.}
\label{tab:pwilcoxon109con}
%\vspace{3mm}
\begin{tabular}{c | c c c c } 
 \hline
 \multicolumn{1}{c|}{} & \multicolumn{2}{|c|}{Pre} & \multicolumn{2}{|c}{Post}  \\
 \hline
  & $\phi_{L_1}$ & $\phi_{L_2}$ & $\phi_{L_1}$ & $\phi_{L_2}$ \\ [0.5ex] 
 \hline\hline
 First & 0.0044(**) & 0.0050(**) & 0.059 & 0.070  \\
 \hline
 Middle & 0.062 & 0.060 & 0.0081(**) & 0.014(*) \\
 \hline
 Last & 0.018(*) & 0.040(*) & 0.0030(**) & 0.0037(**) \\
 \hline
\end{tabular}
\end{center}
\scriptsize{(*) = significant at the $5\%$ significance level. (**) = significant at the $1\%$ significance level. (***) = significant at the $0.1\%$ significance level.}
\end{table}

However, the test results violate the expert's observations: both pre- and
post-surgery data are asymmetric at the first frame and the pre-surgery data
is more asymmetric at this frame. We then compute the $\phi_{L_1}$ and $\phi_{L_2}$
with different weights for the solos, where the weights for the solos
are the doubled unit lengths ($w_j=2$ in (\ref{eq:L1score}) and (\ref{eq:L2score}))
while we still keep the weights for landmark pairs to
be unit length ($w_j=1$ in (\ref{eq:L1score}) and (\ref{eq:L2score})).
The test results from Mann-Whitney U tests are displayed in Table \ref{tab:pwilcoxonprepost}.
$H_0$ is rejected for both groups at the first frame. Further, the $p$-values for
pre-surgery group is more extreme, which indicates that the pre-surgery data
are more asymmetric. This findings match the expert's opinion. Note that
changing the weights in this way do not affect the results for compare-then-combine approach.

\begin{table}[!t]
\begin{center}
\captionsetup{width=1.0\textwidth}
\caption{$p$-values obtained from one-sided two-sample Mann-Whitney U tests using $\phi_{L_1}$ in (equation (\ref{eq:L1score})) and $\phi_{L_2}$ in (\ref{eq:L2score}), where $w_j=1$ for pairs and $w_j=2$ for solos, to compare between Smile Data 2 subjects with control subjects from the Smile Data 1.}
\label{tab:pwilcoxonprepost}
%\vspace{3mm}
\begin{tabular}{c | c c c c } 
 \hline
 \multicolumn{1}{c|}{} & \multicolumn{2}{|c|}{Pre} & \multicolumn{2}{|c}{Post}  \\
 \hline
  & $\phi_{L_1}$ & $\phi_{L_2}$ & $\phi_{L_1}$ & $\phi_{L_2}$ \\ [0.5ex] 
 \hline\hline
 First & 0.0047(**) & 0.0035(**) & 0.026(*) & 0.018(*)  \\
 \hline
 Middle & 0.073 & 0.075 & 0.0055(**) & 0.0073(**) \\
 \hline
 Last & 0.015(*) & 0.042(*) & 0.0018(**) & 0.0026(**) \\
 \hline
\end{tabular}
\end{center}
\scriptsize{(*) = significant at the $5\%$ significance level. (**) = significant at the $1\%$ significance level. (***) = significant at the $0.1\%$ significance level.}
\end{table}

The patterns of the $p$-values shown in Table \ref{tab:as1resultjawsmilepre}, \ref{tab:pwilcoxon109con} and
\ref{tab:pwilcoxonprepost} are the same: the pre-surgery data has smaller $p$-values
at the first frame, whereas the post-surgery data has more extreme $p$-values at the 
middle and last frames.

\subsubsection{Application of the Compare-then-combine Approach}\label{sec:compare-combine-Data1-2}
%\textbf{ approach for Smile Data 1 and Smile Data 2}
\textbf{Feature selection}
The method 
described in Section \ref{sec:test:uit} is used to select the landmarks 
which possess significant asymmetry information for the pre- and post-surgery data
when compare with the normal subjects. 
$v_j$ is selected to be the two-sample t-statistics (equation (\ref{eq:v-stat})).
$V$ defined in (\ref{eq:max}) is computed and its critical value 
at $5\%$ significance level is determined via bootstrap. 
The total number of bootstrap iterations used is 10000.

The central pair, landmarks (3, 5), possesses t-values 
which exceed the critical values of $V$ in most of the cases for both pre- and post-surgery data.
The solo landmark 9 also corresponds to t-values exceed the critical values of
$V$ in some cases. Most of the extreme p-values obtained from Mann-Whitney U
tests pertain to the central pair 3, 5 and two solos, 4 and 9, where $\phi_{L_1}$
and $\phi_{L_2}$ are used.
Landmark pair (3, 5) is the same as landmark pair (6, 8) and solo landmarks 4 and 9 are the
same as landmarks 7 and 19 in Figure \ref{cleftpatient}, 
so we recover the `Y'-shape as in Section \ref{sec:smile}.
% and the solo landmarks 4 and 9

\textbf{Meta-analysis} 
The Fisher's and Pearson's methods given in (\ref{eq:test-Fisher}) and (\ref{eq:test-Pearson})
are executed to compare the subjects from Smile Data 2
with control subjects from the Smile Data 1. $P_R$ defined in equation
(\ref{eq:test-Fisher}) is computed at the three frames separately and compared with the
tail of $\chi^2_{28}$ (as $J=14$). 

\textbf{Pre-surgery data} Table \ref{tab:preconmeta-result} shows the results at the three
frames which are obtained by comparing pre-surgery data with control subjects.
Comparing with Table \ref{tab:as1resultjawsmilepre} and \ref{tab:pwilcoxonprepost},
$H_0$ is also rejected at the middle frame for the Fisher's method and the
$p$-values are more extreme at the first and last frames.
The reason could be that the means of some asymmetry features are different
for the two groups. However, these features have small magnitudes, so they do not contribute
significantly towards the values of composite scores. Consequently, they do not affect the
test results on the composite scores, but impact the outcomes of Fisher's method.
Nevertheless,
for the Pearson's method, the results are the same as in Table \ref{tab:as1resultjawsmilepre}
and \ref{tab:pwilcoxonprepost} and the magnitudes of $p$-values are similar as well.

\begin{table*}[t!]
    \begin{center}
        \captionsetup{width=1.0\textwidth}
    \caption{The results of $P_R$ and $p$-values obtained from Fisher's method and Pearson's method at all three frames. Pre-surgery data from the Smile Data 2 is compared with the control subjects from the Smile Data 1 here.}
    \label{tab:preconmeta-result}
    %\vspace{2mm}
    \begin{tabular}{c|c c c c c }
    \hline
         & $P_L$ & $P_R$ & $Q$ & Fisher's method & Pearson's method \\
         \hline\hline
        First frame & 9.54 & 67.13 & 67.13 & $4.63 \times 10^{-5}$(***) & 0.003(**) \\
         \hline
        Middle frame & 22.25 & 46.44 & 46.44 & 0.016(*) & 0.11 \\
         \hline
        Last frame & 17.16 & 51.90 & 51.90 & 0.0039(**) & 0.038(*) \\
         \hline
    \end{tabular}
    \end{center}
\scriptsize{(*) = significant at the $5\%$ significance level. (**) = significant at the $1\%$ significance level. (***) = significant at the $0.1\%$ significance level.}
\end{table*}

\textbf{Post-surgery data} Table \ref{tab:postconmeta-result} shows the results obtained by comparing 
post-surgery data and control subjects. We obtain the same test results as in
Table \ref{tab:as1resultjawsmilepre} and \ref{tab:pwilcoxonprepost}. The magnitudes
of $p$-values for Pearson's method are similar with the $p$-values obtained
for $\phi_{L_1}$ and $\phi_{L_2}$ at the middle and last frames, whereas the
$p$-value is decreased at the first frame for Pearson's method. For Fisher's method,
the $p$-values are more extreme. 
%(\textbf{\textcolor{red}{We can merge Table 14 and 15 together.}})

\begin{table*}[t!]
    \begin{center}
        \captionsetup{width=1.0\textwidth}
    \caption{The results of $P_R$ and $p$-values obtained from Fisher's method and Pearson's method at all three frames. Post-surgery data from the Smile Data 2 is compared with the control subjects from the Smile Data 1 here.}
    \label{tab:postconmeta-result}
    %\vspace{2mm}
    \begin{tabular}{c|c c c c c }
    \hline
         & $P_L$ & $P_R$ & $Q$ & Fisher's method & Pearson's method \\
         \hline\hline
        First frame & 14.26 & 67.61 & 67.61 & $3.98 \times 10^{-5}$(***) & 0.003(**) \\
         \hline
        Middle frame & 10.83 & 70.64 & 70.64 & $1.51 \times 10^{-5}$(***) & 0.004(**) \\
         \hline
        Last frame & 13.64 & 63.86 & 63.86 & 0.00013(***) & 0.009(**) \\
         \hline
    \end{tabular}
    \end{center}
\scriptsize{(*) = significant at the $5\%$ significance level. (**) = significant at the $1\%$ significance level. (***) = significant at the $0.1\%$ significance level.}
\end{table*}

%\textbf{Compare-then-combine approach using (\ref{eq:test-Fisher}) and (\ref{eq:test-Pearson}):}
%\textbf{\textcolor{red}{A subsection or heading?}}
\subsection{Summary of the Analyses}
We end the section by summarizing our findings on the orthognathic surgery data:
\begin{itemize}
    \item[(a)] The asymmetries are increased after the surgery at middle and last frames.
    \item[(b)] The subjects contained in the Smile Data 2 are more asymmetric than the control subjects from the Smile Data 1, except for the pre-surgery data at the middle frame.
    \item[(c)] The central pair and solo landmarks, i.e. the `Y'-shape in Figure \ref{cleftpatient}, carry important asymmetry information when we compare the subjects in Smile Data 2 with control subjects from Smile Data 1.
\end{itemize}

%\section{To be moved}
%\subsection{Applications}
%We apply the Fisher's and Pearson's method described above to the \textcolor{blue}{two smile datasets}. The two-sample univariate tests are chosen as the Mann-Whitney U tests \textcolor{red}{in comparing
%cleft subjects and control subjects and comparing subjects who have taken orthognathic surgery with control
%subjects (\textbf{redundant})}. The paired univariate tests are chosen as the Wilcoxon signed-rank tests 
%\textcolor{red}{in comparing pre-
%and post-surgery data for the subjects who have taken orthognathic surgery (\textbf{redundant})}.

%For the cleft lip data, we have $J=35$, whereas for the orthognathic surgery data, $J=18$ if we use the 13 landmarks
%at the upper lip region and $J=14$ if we use the 10 landmarks at the lip. \textcolor{blue}{See Section \ref{sec:smile}
%and \ref{sec:jaw} for details.}

%\subsubsection{Analyses of the Cleft Lip Data}

%\subsubsection{Analyses of the Orthognathic Surgery Data}
%\textbf{Effect of the surgery:}
%\textbf{Comparing with the control subjects:}

\section{Discussion}\label{sec:discussion}

%Discussion should contain the following:
We have presented our asymmetry measures for the registered Smile Data. But these asymmetry measures
are available to any registered dataset. 
%but these can be applied after registering the data as described in Section . 
Further, we have used only a size-and-shape framework but the composite measures can be extended easily to include for example, similarity shape, since
%since \textcolor{red}{the hybrid Procrustes analysis method can be used to reduce the similarity shapes to the cases of size-and-shape. Hence,} 
once we have registered the shapes, the asymmetry measures are applicable after allowing for scaling.

We have used absolute values ($a_j$) on coordinate elementary features as in
equation (\ref{absd}) which is of main interest if one is looking for the extent
of asymmetry as in our cases for the illustrative smile examples. However, there maybe some other applications
where raw $d_j$ may be important than $a_j$, such as when analyzing the trajectories of the landmarks; our work easily applies to that situation.
 
In this paper we have concentrated on the illustrative smile examples with three fixed frames and our future statistical work will be extended to dynamic shape analysis which will allow us to study trajectories of the smile. 
%Some initial work for a single group of a smile data has been introduced in \citet{kvm2018} and \citet{bookstein2024}. Some previous works have been done as well, see, for example \citet{morris2000}, \citet{kent2000} and \citet{kent2001}, where the objects after registration are projected to the Procrustes tangent shape space (see Chapter 4 of \citet{drydenmardia2016}) at first, then growth models are fitted.
%\textcolor{red}{We have concentrated on the cleft lip subjects and orthognathic surgery subjects but there are other medical applications such as in facial reconstruction related to trauma surgery where the lips form significant area for surgery.}

In our case, the number of landmarks $K$ is small but when $K$ is large we can study the distribution of asymmetry measure $\phi_{L_2}$ under isotropic normality and various tests can be then performed. The work leads to understanding for the sum of weighted $\chi^2$-distributions. \citet{chisq2025arxiv} studies this approach.

We have used mostly univariate methods except for the Hotelling's $T^2$ in Section \ref{sec:smile}, though in our case, the sample size is less than the dimension of the elementary feature vector so it is not appropriate as Hotelling's $T^2$ is singular. Even in the case when the sample size is large enough,
the one-sided alternative seems natural according to equation (\ref{originH0H1}).
%One-sided multivariate tests have been studied in, for example, \citet{silvapulle19952}, \citet{follmann1996}, \citet{davidov2013} and \citet{arboretti2021}.
Such cases are also part of our future investigation.

We note that the final aim of our work is to provide in future a smile index based on
these asymmetry measures to be used as a standard index in clinics.

%\textcolor{blue}{\citet{chisq2025arxiv} uses another more illustrative way to analyze the 
%smile data, where an isotropic Gaussian
%model is assumed over the landmarks. }

%Hence, the paper examines The paper also extends to the cases where the degrees of freedom
%for $\chi^2$-distribution is large, i.e. when $K_P$ or $K_S$ is large,
%so that the $\chi^2$ can be approximated by a
%normal distribution.

%We can also define the \textit{signed} version of the elementary feature vector $\bm d$ and give it the name $\bm d^{\text{sign}}$. The \textit{signed coordinatewise elementary features} can be defined as
%\begin{align}
 % d^{\text{sign}}[(k_L,k_R),1] &= X[k_L,1]+X[k_R,1] \label{eq:unsignpair1}\\
 % d^{\text{sign}}[(k_L,k_R),m] &= X[k_L,m]-X[k_R,m], \quad m=2, \ldots, M. \label{eq:unsignpair2}.
%\end{align}
\nil{The motivation for the composite score is to ``combine strength'' from
the elementary features, since we expect them all to be larger (on
average) for the more asymmetric group.}
%Rational for multivariate testing. -- still need to incorporate XW draft.
%In this setting we work with the elementary feature vectors directly rather than summarizing them with a scalar composite score.

\section*{Data Availability Statement}
The codes and data for this paper are available at the following website: \url{https://github.com/XW-2025-hub/JRSS-C-Supplementary-Materials}.

\section*{Acknowledgments}
The authors would like to thank the School of Mathematics, and the School of Dentistry  University of Leeds for supporting the smile project; we are also grateful to Fred Bookstein for collaborating with this project. The authors would also like to thank Sofya Titarenko for her helpful comments. Kanti Mardia acknowledges the
Leverhulme Trust for the Emeritus Fellowship. 
   % \textbf{To be done.}

\nil{\begin{appendices}
\end{appendices}}

\bibliographystyle{rss}
\bibliography{references}

\end{document}